\documentclass[aps,prb,showpacs,twocolumn,superscriptaddress]{revtex4-1}
\usepackage{bm,color,amsmath,amssymb,mathrsfs,latexsym,graphicx,psfrag,accents,float}
\pdfoutput=1

\newcommand{\zz}{{\zeta}_0}
\newcommand{\zo}{{\zeta}_1}
\newcommand{\zt}{{\zeta}_2}

\newcommand{\go}{\gamma_{\sma{1}}}
\newcommand{\gt}{\gamma_{\sma{2}}}
\newcommand{\gth}{\gamma_{\sma{3}}}
\newcommand{\gfo}{\gamma_{\sma{4}}}
\newcommand{\gfi}{\gamma_{\sma{5}}}
\newcommand{\gsix}{\gamma_{\sma{6}}}

\newcommand{\gtj}{\gamma_{\sma{2j}}}
\newcommand{\gtjpo}{\gamma_{\sma{2j+1}}}
\newcommand{\gtjpt}{\gamma_{\sma{2j+2}}}
\newcommand{\gtjmo}{\gamma_{\sma{2j-1}}}
\newcommand{\gtL}{\gamma_{\sma{2L}}}

\newcommand{\gtLmo}{\gamma_{\sma{2L-1}}}
\newcommand{\gtLmt}{\gamma_{\sma{2L-2}}}
\newcommand{\gtLmth}{\gamma_{\sma{2L-3}}}
\newcommand{\gtLmfo}{\gamma_{\sma{2L-4}}}
\newcommand{\gtLmfi}{\gamma_{\sma{2L-5}}}

\newcommand{\calkxyzalpha}{\calk^{\sma{(\alpha)}}_{\sma{XYZ}}}
\newcommand{\calhxyz}{\calh_{\sma{XYZ}}}
\newcommand{\calkxyz}{\calk_{\sma{XYZ}}}
\newcommand{\calgxyz}{\calg_{\sma{XYZ}}}
\newcommand{\hxyz}{H_{\sma{XYZ}}}
\newcommand{\hno}{H_{\sma{(N,1)}}}
\newcommand{\het}{H_{\sma{(8,2)}}}
\newcommand{\pno}{P_{\sma{(N,1)}}}
\newcommand{\hnn}{H_{\sma{(N,n)}}}

\newcommand{\pnn}{P_{\sma{(N,n)}}}
\newcommand{\gno}{\calg_{\sma{(N,1)}}}
\newcommand{\pft}{P_{\sma{(4,2)}}}

\newcommand{\kal}{\calk^{\sma{(\alpha)}}}

\newcommand{\ins}[1]{\;\;\;\;\text{#1}\;\;\;\;}

\newcommand{\calb}{{\cal B}}

\newcommand{\cald}{{\cal D}}

\newcommand{\calg}{{\cal G}}
\newcommand{\calh}{{\cal H}}

\newcommand{\calk}{{\cal K}}
\newcommand{\call}{{\cal L}}
\newcommand{\calm}{{\cal M}}

\newcommand{\calo}{{\cal O}}

\newcommand{\calq}{{\cal Q}}

\newcommand{\cals}{{\cal S}}
\newcommand{\calt}{{\cal T}}

\newcommand{\calv}{{\cal V}}

\newcommand{\calx}{{\cal X}}
\newcommand{\caly}{{\cal Y}}
\newcommand{\calz}{{\cal Z}}

\newcommand{\noi}[1]{\noindent (#1)}
\newcommand{\imp}{\;\;\Rightarrow\;\;}
\newcommand{\mo}{\text{-}1}

\newcommand{\oneover}[1]{\tfrac{1}{#1}}

\newcommand{\braket}[2]{\big\langle #1 \big| #2 \big\rangle}
\newcommand{\bra}[1]{\big\langle#1\big|}
\newcommand{\ket}[1]{\big|#1\big\rangle}

\newcommand{\half}{\tfrac{1}{2} }

\newcommand{\bea}{\begin{eqnarray}}
\newcommand{\enea}{\end{eqnarray}}
\newcommand{\beq}{\begin{equation}}
\newcommand{\eneq}{\end{equation}}

\newcommand{\pdg}[1]{{#1}^{\phantom{\dagger}}}

\newcommand{\lin}{\notag \\}

\newcommand{\eq}{=&\;}
\newcommand{\ab}{\alpha\beta}

\newcommand{\bpm}{\begin{pmatrix}}
\newcommand{\epm}{\end{pmatrix}}

\newcommand{\bal}{\begin{align}}
\newcommand{\eal}{\end{align}}

\newcommand{\si}{\;\text{sin}\,}

\newcommand{\dg}[1]{#1^{\scriptstyle{\dagger}}}

\newcommand{\sma}[1]{\scriptscriptstyle{#1}}

\newcommand{\Z}{\mathbb{Z}}

\newcommand{\qed}{\nobreak \ifvmode \relax \else
      \ifdim\lastskip<1.5em \hskip-\lastskip
      \hskip1.5em plus0em minus0.5em \fi \nobreak
      \vrule height0.75em width0.5em depth0.25em\fi}

\begin{document}
\title{Parafermionic phases with symmetry-breaking and topological order} 
 \author{A. Alexandradinata} \affiliation{Department of Physics, Princeton University, Princeton,
  NJ 08544, USA} 
  \author{N. Regnault} \affiliation{Department of Physics, Princeton University, Princeton,
  NJ 08544, USA} \affiliation{Laboratoire Pierre Aigrain, Ecole Normale Sup\'erieure-PSL Research
University, CNRS, Universit\'e Pierre et Marie Curie-Sorbonne Universit\'es,
Universit\'e Paris Diderot-Sorbonne Paris Cit\'e, 24 rue Lhomond, 75231
Paris Cedex 05, France}		
  \author{Chen Fang} \affiliation{Department of Physics, Princeton University, Princeton,
  NJ 08544, USA} 
   \affiliation{Department of Physics, University of Illinois, Urbana IL 61801, USA}
   \affiliation{Micro and Nanotechnology Laboratory, University of Illinois, 208 N. Wright Street, Urbana IL 61801, USA}
\author{Matthew J. Gilbert} \affiliation{Micro and Nanotechnology Laboratory, University of Illinois, 208 N. Wright Street, Urbana IL 61801, USA}
  \affiliation{Department of Electrical and Computer Engineering, University of Illinois, Urbana IL 61801, USA}
  \author{B. Andrei Bernevig} \affiliation{Department of Physics, Princeton University, Princeton,
  NJ 08544, USA}


\begin{abstract}
Parafermions are the simplest generalizations of Majorana fermions that realize topological order. We propose a less restrictive notion of topological order in 1D {open} chains, which generalizes the seminal work by Fendley [J. Stat. Mech., P11020 (2012)]. The first essential property is that the groundstates are mutually indistinguishable by local, symmetric probes, and the second is a generalized notion of zero edge modes which cyclically permute the groundstates. These two properties are shown to be topologically robust, and applicable to a wider family of topologically-ordered Hamiltonians than has been previously considered. An an application of these edge modes, we formulate a new notion of twisted boundary conditions on a closed chain, which guarantees that the closed-chain groundstate is topological, i.e., it originates from the topological manifold of degenerate states on the open chain. Finally, we generalize these ideas to describe symmetry-breaking phases with a \emph{parafermionic} order parameter. These exotic phases are condensates of parafermion multiplets, which generalizes Cooper pairing in superconductors. The stability of these condensates are investigated on both open and closed chains.
\end{abstract}
\date{\today}


\maketitle

\section{Introduction}

Parafermions are the simplest generalizations of Majorana fermions\cite{Majorana} that realize topological order.\cite{fendley2012} One signature of topological order is the manifestation of non-Abelian anyons\cite{moore1991,read2000,ivanov2001} on a 1D open chain,\cite{kitaev2001} which has applications in topological quantum computation.\cite{kitaev2003,nakakreview2008,alicea2011} For practical computation, it is desirable that the anyons robustly survive any generic local perturbation. Following a seminal work by Fendley,\cite{fendley2012} there is strong evidence that topological order is indeed stable over a family of $\Z_N$-symmetric Hamiltonians with broken parity and time-reversal symmetries. On an open chain, these so-called chiral Hamiltonians\cite{ostlund,huse1981,huse1983,howes1983,albertini,auyang} satisfy an additional symmetry that renders their \emph{entire} spectra $N$-fold degenerate.\cite{fendley2012} We propose that this symmetry is sufficient but not necessary for a topologically ordered phase, since topological order describes the low-energy subspace, rather than the entire range of excitations. By relaxing this symmetry constraint, we are led to consider a wider family of topologically-ordered Hamiltonians than has been considered in Ref.\ \onlinecite{fendley2012}; these include the non-chiral Hamiltonian which is dual to the widely-studied ferromagnetic clock model.\cite{potts,elitzur} Despite the loss of degeneracy in the high-energy subspace, the groundstate space remains $N$-fold degenerate to superpolynomial accuracy in the system size. We are able to demonstrate a stronger statement, that the groundstates are mutually indistinguishable by \emph{all} local, symmetry-preserving operators, of which the Hamiltonian is but one case in point. \\

\begin{figure}
\centering
\includegraphics[width=8.3 cm]{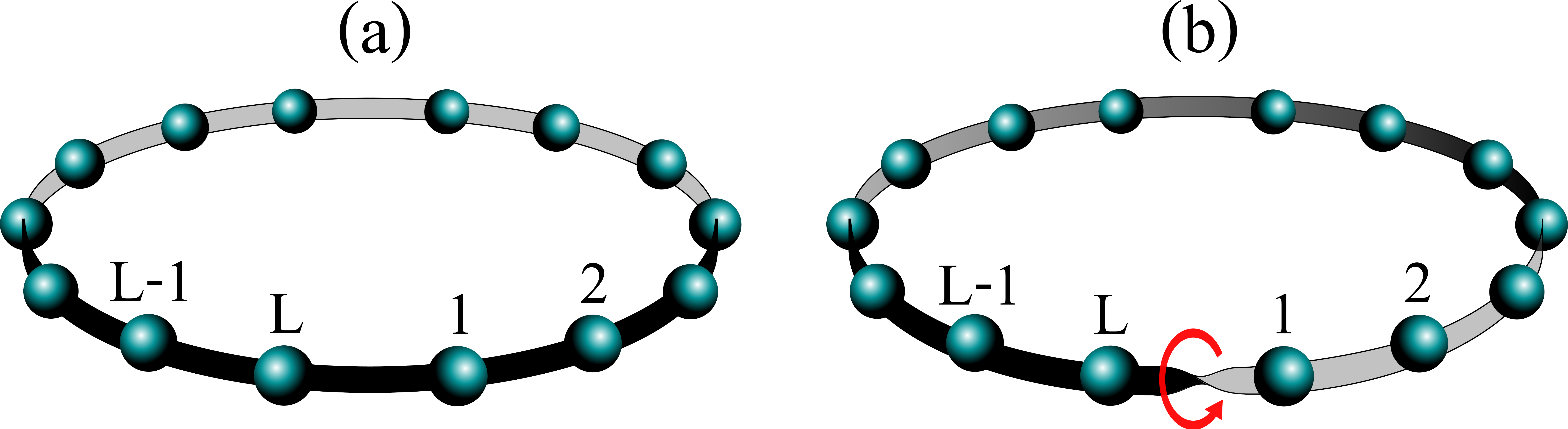}
\caption{ We formulate a new notion of twisted boundary conditions on the closed chain. A topological groundstate is especially sensitive to a change in boundary conditions (schematically as (a) to (b)). Such a twist results in a different groundstate with a distinct $\Z_N$ charge. }\label{fig:moebius}
\end{figure}

The nontrivial groundstate degeneracy on an open chain is intimately connected to the presence of edge operators which permute the various groundstates. The best-known examples include the dangling Majorana fermion on the edge of the Kitaev chain,\cite{kitaev2001,semenoff2007} as well as various other proposals to realize Majorana bound states.\cite{fu2008,*fukanejosephson,lutchyn2010,oreg2010,potter2010,stanescu,mourik,das2012,stevan} A related signature is the nontrivial degeneracy in the entanglement spectrum, where a real-space bipartition introduces a virtual edge.\cite{turner2011,motruk2013} To precisely formulate this `dangling parafermion', we introduce the notion of a pair of operators ($\calo_l,\calo_r$), which localize on left and right edges of an open chain. Each of the pair permutes the groundstates cyclically, but does not necessarily commute with the full Hamiltonian, thus differing from past proposals.\cite{fendley2012,kells2014} Our generalized edge modes are found to generate a non-commutative algebra within the groundstate space, which for $N=2$ is the well-known Clifford algebra. \\

The topological degeneracy that we attribute to `dangling parafermions' is generically lost on a closed chain without edges.\cite{motruk2013,bondesan2013} One might therefore ask: is the non-degenerate, closed-chain groundstate continuously connected to the open-chain topological manifold? In this paper, we set out to answer a related question: what is the closed-chain Hamiltonian ($\calh$) that is \emph{exactly} minimized by a state in the open-chain manifold? We find that $\calh$ is obtained by utilizing $\calq \sim \dg{\calo}_l\pdg{\calo}_r$ as an inter-edge coupling to close the chain. Since $\calq$ breaks translational-invariance, our proposed Hamiltonian differs from conventional methods\cite{fidkowski2011b,zaletel2014} to close the chain. $\calq$ may be interpreted as a topological order parameter, in analogy with the local order parameter ($\hat{O}$) of traditional broken-symmetry phases. Traditionally, by applying a local, symmetry-breaking `field' $\lambda\hat{O}+h.c.$, one picks out a broken-symmetry groundstate that depends on the phase factor $\lambda$. Analogously, by applying $\lambda \calq+h.c.$ to a topological manifold on an open chain, one selects a closed-chain groundstate that originates from this manifold. By exploring the space of $\lambda$, we are able to select a different state from the topological manifold, with a different $\Z_N$ charge; this variation of $\lambda$ is analogous to twisting the boundary conditions in integer quantum Hall systems.\cite{laughlin1981}  \\


While topologically-ordered parafermions naturally generalize the $\Z_2$ Kitaev chain,\cite{kitaev2001} some parafermionic phases have no Majorana analog. Indeed, $\Z_{\sma{N}}$ parafermions with $N \geq 4$ can also exhibit symmetry-breaking, sometimes even in conjunction with topological order.\cite{bondesan2013,motruk2013} Symmetry-breaking due to a parafermionic order parameter ($\caly$) is nonlocal, and thus fundamentally different from traditional symmetry-breaking by a local order parameter. Expanding on an interpretation in Ref.\ \onlinecite{motruk2013}, we identify these exotic phases as condensates of `Cooper multiplets', which generalize Cooper pairing in superconductors. A major advance of Ref.\ \onlinecite{bondesan2013} is the construction of a class of frustration-free Hamiltonians $\hnn$, which realize all possible distinct phases. For these fine-tuned Hamiltonian, the exact form of $\caly$ is now known (when there is symmetry-breaking), as is the exact form of the zero edge modes (when there is topological order). On the other hand, the stability of these phases beyond the frustration-free limit $\hnn$ has been posed as a conjecture.\cite{bondesan2013} This lack of quantitative understanding has led to controversy, e.g., with regard to the correct groundstate degeneracy on a closed chain in the presence of symmetry-breaking.\cite{bondesan2013,motruk2013} One goal of this paper is to settle this controversy and prove this conjecture. \\

The outline of our paper: we review the simplest models that realize topological order in Sec.\ \ref{sec:notation}, and describe the strict notion of a zero edge mode on an open chain.\cite{fendley2012}. After arguing that this strict notion is sufficient but not necessary for topological order, we then formulate two essential properties of topological phases on an open chain. As described in Sec.\ \ref{sec:localindis}, the first is that the groundstate space is mutually indistinguishable to symmetric, local probes. The second property is a generalized notion of zero edge modes, which we show to have several interesting applications in Sec.\ \ref{sec:zeromodes}. One application is the construction of a topological order parameter ($\calq$), which lends us a new notion of twisted boundary conditions on a closed chain. This notion is formulated in Sec.\ \ref{app:detectTOonring}; a critical comparison is made with the conventional method of twisting. In Sec.\ \ref{sec:pfsymmbreak}, we generalize these concepts to exotic phases with parafermionic order parameters. The stability of these phases is investigated on both open and closed chains. In the last Sec.\ \ref{sec:discussion}, we discuss the possible generalizations of our work.


\section{Review of parafermionic phases with topological order} \label{sec:notation}

\begin{figure}[H]
\centering
\includegraphics[width=8.3 cm]{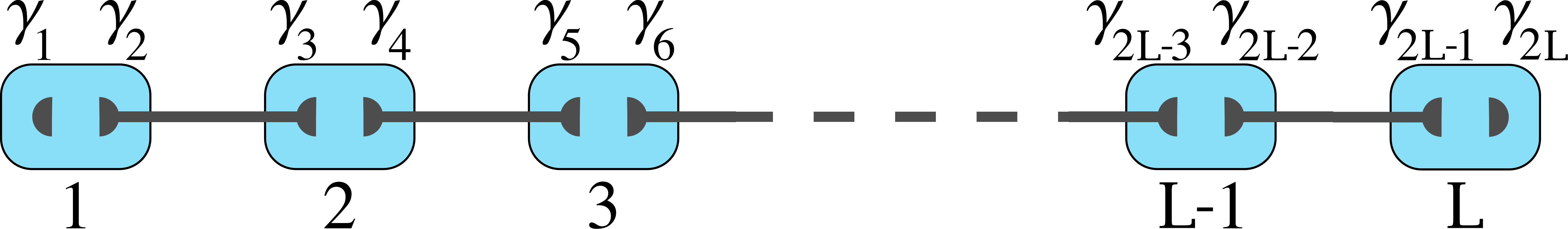}
\caption{Parafermion chain of $L$ sites. Each site (blue circle) is split into two subcells. }\label{fig:parafermionchain}
\end{figure}

Parafermions generalize the Majorana algebra through 
\begin{align} \label{parafermionalgebra}
\gamma_{\sma{j}}^N=1,\;\;\dg{\gamma}_{\sma{j}} = \gamma_{\sma{j}}^{N-1} \ins{and} \pdg{\gamma}_{\sma{j}} \pdg{\gamma}_{\sma{l}} = \omega^{\text{sgn}[l-j]}\pdg{\gamma}_{\sma{l}}\pdg{\gamma}_{\sma{j}},
\end{align}
for $l\neq j$, and $\omega=e^{i2\pi/N}$. Here, we have denoted $\gamma_{j}$ as an operator acting on subcell $j \in \{1,2,\ldots,2L\}$ on a chain of $L$ sites, in the Hilbert space $(\mathbb{C}^N)^{\otimes L}$. Our convention is to pair subcells $2j-1$ and $2j$ into a single site $j$, as illustrated in Fig.\ \ref{fig:parafermionchain}. The simplest, fixed-point\cite{Xie2011} Hamiltonians having topological order are dimerized as
\begin{align} \label{pfHamn1}
\hno =&\; -J \sum_{j=1}^{L-1} \sum_{\beta=0}^{N-1}(\omega^{(N-1)/2} \dg{\gamma}_{\sma{2j+1}}\pdg{\gamma}_{\sma{2j}})^{\beta},
\end{align}
where each `even' parafermion (on an even subcell) is coupled only to one `odd' parafermion on the adjacent site; these couplings are schematically drawn as black lines in Fig.\ \ref{fig:parafermionchain}. On an open chain, there is then a dangling parafermion on each end which is uncoupled from the bulk of the chain. For $N=2$, Eq.\ (\ref{pfHamn1}) is the well-known Kitaev chain with a dangling Majorana mode on each end.\cite{kitaev2001} $\hno$ has a $\Z_N$ symmetry which is generated by the string operator
\begin{align} \label{Zngenerator}
Q = \prod_{j=1}^L (\omega^{(1-N)/2}\dg{\gamma}_{\sma{2j-1}} \pdg{\gamma}_{\sma{2j}}),
\end{align}
i.e., $[\hno,Q]=0$. $Q$ generalizes the fermion parity of Majorana systems through $Q^N=I$. We refer to an eigenstate of $Q$ with eigenvalue $\omega^{\alpha}$ as belonging to the charge sector $\alpha \in \{0,1,2,\ldots, N-1\} \equiv \Z_N $. Each dangling parafermion  $\calo \in \{\gamma_{\sma{1}},\pdg{\gamma}_{\sma{2L}}\}$ is a localized unitary operator satisfying
\begin{align} \label{Fzero}
[\hno,\calo]=0 \ins{and} \calo Q = \omega Q \calo.
\end{align}
These symmetry relations imply that the entire spectrum is $N$-fold degenerate, where each $N$-multiplet comprises a state in each charge sector. Fendley defines any operator satisfying Eq.\ (\ref{Fzero}) as a zero edge mode.\cite{fendley2012} We refer to $\hno$ and its equivalence class as purely-topological, in the sense that its groundstate space has topological order, but is not symmetry-broken by a parafermionic order parameter; cf. Sec.\ \ref{sec:pfsymmbreak}. As remarked before, $\hno$ belongs to a special class of fine-tuned Hamiltonians that is frustration-free. To clarify, let us decompose $\hno$ into a sum of two-site operators:  $h_{\sma{j+1,j}}=\sum_{\sma{\beta=0}}^{\sma{N-1}}(\omega^{\sma{(N-1)/2}} \gamma^{\sma{\dagger}}_{\sma{2j+1}}\pdg{\gamma}_{\sma{2j}})^{\sma{\beta}}$. By frustration-free, we mean that each of $\{h_{\sma{j+1,j}}\}$ is a mutually-commuting projection, and is individually minimized by each groundstate. $h_{\sma{j+1,j}}$ has an intuitive interpretation in the clock representation of parafermion operators, which we now describe.  The clock representation is obtained by the generalized Jordan-Wigner transformation:\cite{fradkin1980}
\begin{align} \label{pfclocktransformation}
\pdg{\gamma}_{\sma{2j-1}}= \sigma_j \prod_{k=1}^{j-1} \tau_k, \ins{and} \pdg{\gamma}_{\sma{2j}} = \omega^{(N-1)/2} \sigma_j \prod_{k=1}^j \tau_k,
\end{align}
where $\sigma_j$ and $\tau_j$ act on site $j$. These clock operators satisfy the algebra
\begin{align}
\sigma_j^N=&\tau_j^N=1,\;\;\dg{\sigma}_j =\sigma_j^{N-1},\;\;\dg{\tau}_j =\tau_j^{N-1}, \lin
&\ins{and} \tau_j\sigma_j\tau_j^{\mo} = \omega^*\sigma_j.
\end{align} 
By this nonlocal transformation, we find that Eq.\ (\ref{pfHamn1}) is dual to the ferromagnetic clock model:\cite{potts}
\begin{align} \label{clockhamn1}
\hno =&\; -J \sum_{j=1}^{L-1} \sum_{\beta=0}^{N-1}(\pdg{\sigma}_j \dg{\sigma}_{j+1})^{\beta},
\end{align}
which has a $\Z_N$ symmetry generated by $Q=\prod_{j=1}^L\tau_j$. To be transparent, let us define an $N$-dimensional basis on each site $j$, satisfying
\begin{align} \label{basisonsite}
\sigma_j\ket{\alpha}_{\sma{j}} = \omega^{\alpha}\ket{\alpha}_{\sma{j}},\;\; \tau_j\ket{\alpha}_{\sma{j}} = \ket{\alpha+1\;\text{mod}\;N}_{\sma{j}},
\end{align}
for $\alpha \in \Z_N$. Each of $h_{\sma{j+1,j}}$ projects to a ferromagnetic alignment of the clock variables on adjacent sites: $|\alpha\rangle_{\sma{j}}\otimes |\alpha\rangle_{\sma{j+1}}$. Thus, the groundstate of $\hno$ exhibits long-range order with respect to the order parameter $\sigma_j$, i.e.,  $\langle \pdg{\sigma}_{\sma{i}}\dg{\sigma}_{\sma{j}} \rangle$ is nonzero for large $|i-j|$. This type of symmetry-breaking by a clock order parameter has a long history.\cite{nachtergaele1996} In this paper, we also describe a different type of symmetry-breaking by a parafermionic order parameter, which is the subject of Sec.\ \ref{sec:pfsymmbreak}.

\begin{figure}[H]
\centering
\includegraphics[width=6 cm]{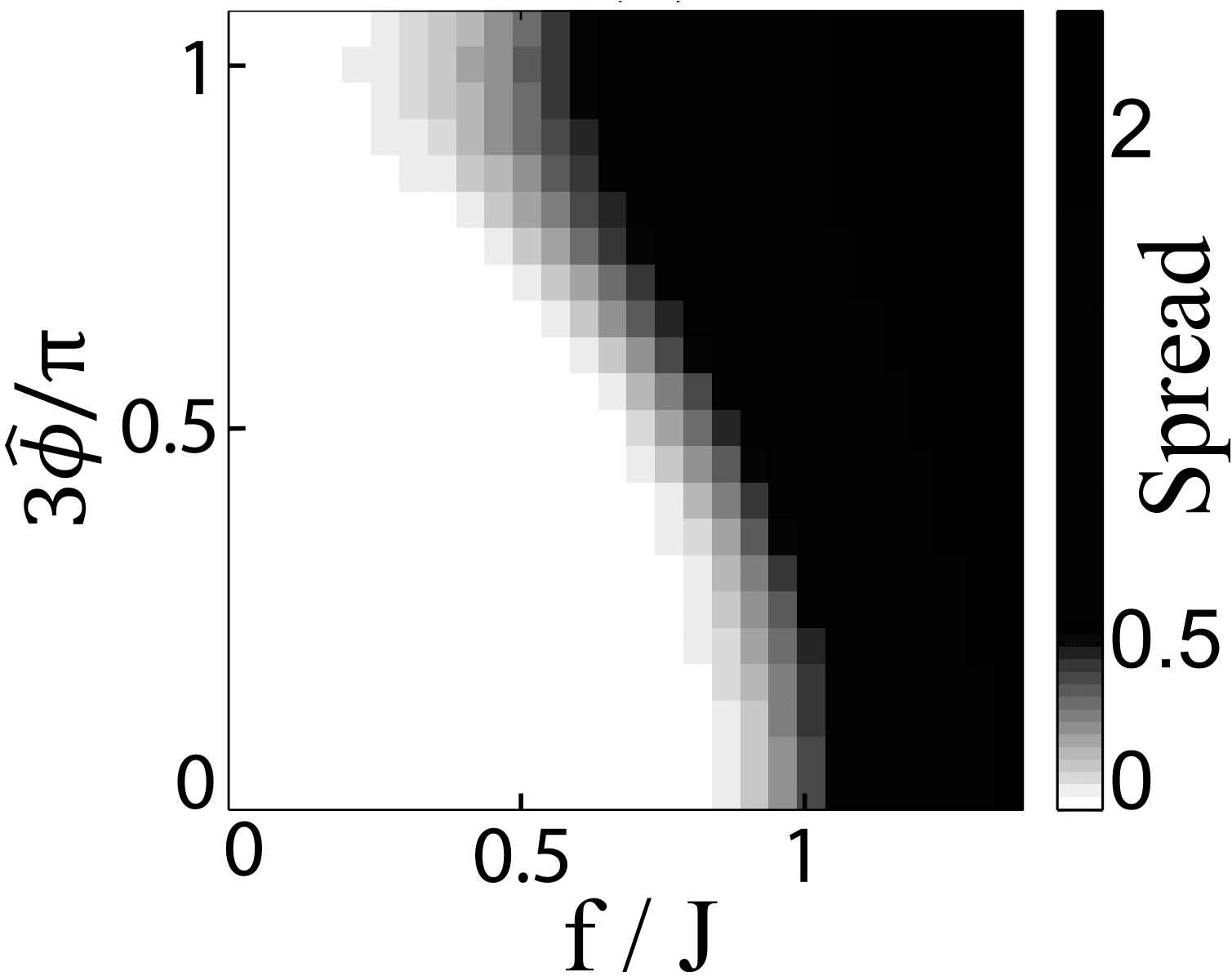}
\caption{Numerical simulation of $\Z_3$ parafermions on a $14$-site chain. We plot the absolute spread in energy between three states $\{|\tilde{\Psi}_{\sma{\alpha}}\rangle\}$, where $|\tilde{\Psi}_{\sma{\alpha}}\rangle$ is the groundstate of the open-chain Hamiltonian\ (\ref{Z3example}) in the charge sector $\alpha \in \Z_3$. }\label{fig:Z3}
\end{figure}

When deviating from the frustration-free $\hno$, it is not obvious if such localized operators ($\calo$) satisfying Eq.\ (\ref{Fzero}) still exist. For illustration, we consider a well-studied deformation with $\Z_3$ symmetry:
\begin{align} \label{Z3example}
\hat{H} =-Je^{i\hat{\phi}}\sum_{j=1}^{L-1}\omega \dg{\gamma}_{\sma{2j+1}}\pdg{\gamma}_{\sma{2j}} -f\sum_{j=1}^L\omega^*\dg{\gamma}_{\sma{2j-1}} \pdg{\gamma}_{\sma{2j}} + h.c.,
\end{align}
which reduces to $H_{\sma{(3,1)}}$ (up to a constant) if we set $f=\hat{\phi}=0$. The $f$-term is frustrating, in the sense that it cannot be minimized simultaneously with the $J$-term. This frustration tends to destabilize the ordered phase, such that localized operators satisfying Eq.\ (\ref{Fzero}) no longer exist where $\hat{\phi}=0$. However, Fendley has nontrivially demonstrated that such edge modes persist if $\hat{\phi}$ is deformed away from $0$;\cite{fendley2012} this is also understood through domain-wall dynamics in the dual clock representation.\cite{jermyn2014} Since both time-reversal and spatial-inversion symmetries are broken when $\hat{\phi} \neq 0$ mod $\pi/3$, the resulting Hamiltonians are called chiral;\cite{ostlund,huse1981,huse1983,howes1983,albertini,auyang} it has been suggested that breaking these symmetries changes the type of topological order, or eliminates it altogether.\cite{fendley2012}  This is peculiar because the non-chiral Hamiltonian\ (\ref{pfHamn1}) at $\hat{\phi}=0$  exhibits the largest spectral gap as a function of $\hat{\phi}$, thus one expects here that the topological properties are most robust. To substantiate this claim, we numerically evaluate the groundstate degeneracy in Fig.\ \ref{fig:Z3}; this degeneracy is shown to be most persistent along the line parametrized by $(f,\hat{\phi}=0)$. Our point of view is that topological order is a property of the groundstate space, and any constraint on the excited states is not necessary, Eq.\ (\ref{Fzero}) being a case in point. This motivates us to formulate two essential characteristics of the groundstate space, which applies to a wider family of Hamiltonians than has been previously considered: (i) a persistent groundstate degeneracy, despite the loss of degeneracy in the excited spectrum  (elaborated in Sec.\ \ref{sec:localindis}), and (ii) the existence of generalized zero edge modes which do not satisfy  Eq.\ (\ref{Fzero}), but nonetheless generate a non-commutative algebra in the groundstate space, as we will show in  Sec.\ \ref{sec:zeromodes}.

\section{Local indistinguishability in purely-topological phases} \label{sec:localindis}

Consider the $\Z_3$, non-chiral Hamiltonian in Eq.\ (\ref{Z3example}) with $\hat{\phi}=0$. With $f=0$ as well, the entire spectrum is three-fold degenerate. With nonzero $f$, it is known that the groundstate splitting is exponentially-small in the system size $L$, while the splittings of the excited triplets scales as a power law.\cite{jermyn2014} This suggests a general property of topologically-ordered systems, that a nontrivial groundstate degeneracy can persist without degeneracy in the excited states. In this section, we confirm this general property, and also formulate a stronger statement: that the groundstates are mutually indistinguishable by \emph{any} quasilocal, symmetric operator ($\calo$). By quasilocal, we mean that each term in $\calo$ has a support that is finite-ranged or decaying faster than any power, i.e., superpolynomially; see App.\ \ref{app:quasilocal} for a precise definition of quasilocality. \\

For open-chain, quasilocal Hamiltonians that are topologically-ordered (without symmetry-breaking), we formulate a notion of local indistinguishability in their groundstate spaces.  We define $P^{\sma{(\alpha)}}$ as the nondegenerate groundstate projection in the charge sector labelled by $\alpha \in  \Z_N$, i.e., the groundstate has eigenvalue $\omega^{\alpha}$ under operation by $Q$. Let $\pno=\sum_{\alpha =0}^{N-1}P^{\sma{(\alpha)}}$, and let $\calo$ denote \emph{any} quasilocal, symmetry-preserving operator, i.e., $[\calo,Q]=0$. In the thermodynamic limit, 
\begin{align} \label{spo}
\pno\, \calo \,\pno = c(\calo)\,\pno,
\end{align}
for some complex number $c$, as we prove in the remainder of this Section. Since the open-chain Hamiltonian $H_{\sma{op}}$ satisfies the same conditions as $\calo$, Eq.\ (\ref{spo}) also implies that the groundstate is $N$-fold degenerate with energy $c(H_{\sma{op}})$.\\

As a first step, we will prove that Eq.\ (\ref{spo}) is satisfied by the frustration-free $\hno$, and then we will extend our result to frustrated Hamiltonians. Indeed, for finite-ranged $\calo$, Eq.\ (\ref{spo}) is satisfied by $\hno$ exactly, without finite-size corrections. This is most conveniently proven in the clock representation\ (\ref{clockhamn1}) of $\hno$. Each of $N$ groundstates is a classical product state that is fully-polarized in the quantum numbers of $\sigma_j$: $|\psi_{\sma{\alpha,1}}\rangle = \bigotimes_{\sma{j=1}}^{\sma{L}} |\alpha\rangle_{\sma{j}}$ for $\alpha \in \Z_N$. The property\ (\ref{spo}) has a simple explanation in the clock representation: local operators have difficulty transforming one polarized state $|\psi\rangle$ into another. To substantiate this intuition, we first define an alternative groundstate basis which diagonalizes $Q$. For $\alpha \in \Z_N$, let
\begin{align} \label{definephi}
\ket{\phi_{\alpha,1}} =\frac{1}{\sqrt{N}}\sum_{\beta=0}^{N-1} {\omega}^{-\alpha \beta} \ket{\psi_{\beta,1}},
\end{align}
such that $Q|\phi_{\sma{\alpha,1}}\rangle = \omega^{\alpha}|\phi_{\sma{\alpha,1}}\rangle$; we refer to $\alpha$ as the $\Z_N$ charge of $|\phi_{\sma{\alpha,1}}\rangle$. Eq.\ (\ref{spo}) amounts to demonstrating that the matrix $\calm^{\sma{(1)}}_{\sma{\ab}} = \langle \phi_{\sma{\alpha,1}}|\,\calo\, |\phi_{\sma{\beta,1}}\rangle$ is proportional to the identity. $[\calo,Q]=0$ implies that the off-diagonal elements vanish. The difference in diagonal elements, $\calm^{\sma{(1)}}_{\sma{\alpha \alpha}}-\calm^{\sma{(1)}}_{\sma{\beta \beta}}$, reduces to a sum of terms like $\langle \psi_{\sma{\mu,1}}|\,\calo\, |\psi_{\sma{\nu,1}}\rangle$ for $\mu \neq \nu$. If $\calo$ has finite range, this quantity vanishes because $\calo$ cannot change the polarization on every site. A variant of this proof was first presented in Ref.\ \onlinecite{hastings2005} to explain the robust groundstate degeneracy of the Ising model. If $\calo$ decays superpolynomially instead, $\calm^{\sma{(1)}}_{\sma{\alpha \alpha}}-\calm^{\sma{(1)}}_{\sma{\beta \beta}}$ has a finite-size correction that we bound in App.\ \ref{app:generalprooflocalindist}. \\

While deviating from $\hno$ introduces quantum fluctuations to the classical states $\{|\psi_{\sma{\alpha,1}}\rangle\}$, these fluctuations have a length scale that is suppressed by a spectral gap, thus preserving the property\ (\ref{spo}) to superpolynomial accuracy in the system size. We are assuming a gapped, quasilocal interpolation $H_s$ between $H_0 \equiv \hno$ and a deformed, but topologically equivalent, Hamiltonian $H_1$; the $\Z_N$ symmetry is preserved throughout $s \in [0,1]$. By gapped, we mean that the spectral gap above the lowest $N$ states remains finite throughout the interpolation -- this implies that the projection $P_s$ to the lowest $N$ states is uniquely defined at each $s$. At this point we do not assume that these $N$ states remain degenerate. These conditions allow us to define a locality- and symmetry-preserving unitary transformation $\calv_s$, that maps the low-energy subspaces as
\begin{align} \label{qacmapping}
P_s = \calv_sP_0\dg{\calv}_s, \ins{with} P_0 \equiv \pno,\; [\calv_s,Q]=0.
\end{align}
By locality-preserving, we mean that if $\calo$ is quasilocal, so is its transformed version $\dg{\calv}_s\calo {\calv}_s$. $\calv_s$ is known as an exact quasi-adiabatic continuation;\cite{hastings2005,hastings2010,brayvi2011} its properties are elaborated in App.\ \ref{app:quasilocal}, and its explicit form is shown in the next Section. Under these conditions, we find that if $P_0$ is locally indistinguishable, so is $P_s$. To prove this, let $\calo$ be quasilocal and symmetry-preserving, then $P_s\calo P_s = \calv_sP_0\dg{\calv}_s\calo \calv_sP_0\dg{\calv}_s$. Since $\dg{\calv}_s\calo \calv_s$ is also quasilocal and symmetry-preserving, we may apply Eq.\ (\ref{spo}) that we have shown to be valid for the frustration-free $P_0$. It follows that $P_s\calo P_s =cP_s$ for some constant $c$, up to superpolynomially-small finite-size corrections that we bound in App.\ \ref{app:generalprooflocalindist}. \\

Finally, we point out that local indistinguishability is a unifying property of many other topologically-ordered groundstates, as we elaborate in Sec.\ \ref{sec:discussion}.

\section{Generalized zero edge modes for purely-topological phases} \label{sec:zeromodes}

The $N$-fold degeneracy on an open chain arises solely from degrees of freedom on the edges. To crystallize this notion, we would like to construct localized edge operators which permute the groundstates cyclically. We already know such a set of operators for the frustration-free $H_0 =\hno$: from\ (\ref{pfclocktransformation}) and\ (\ref{definephi}), it is simple to deduce that $\bra{\phi_{\sma{\alpha,1}}}\gamma_{\sma{1}}\ket{\phi_{\sma{\alpha+1 \,\text{mod}\,N,1}}}=1$ and $\bra{\phi_{\sma{\alpha,1}}}\pdg{\gamma}_{\sma{2L}}\ket{\phi_{\sma{\alpha+1\,\text{mod}\,N,1}}}=\omega^{\sma{(N+1)/2+\alpha}}.$  Given this information, we can construct similar operators for any Hamiltonian $H_s$ that is connected to $H_0$ by a gapped, symmetry-preserving interpolation. This will be accomplished by quasi-adiabatic continuation, which is known to preserve the matrix elements of operators within the groundstate space.\cite{hastings2005} In more detail, we decompose the groundstate space of $H_s$ as $P_s=\sum_{\sma{\alpha=0}}^{\sma{N-1}}\ket{\phi^s_{\sma{\alpha,1}}}\bra{\phi^s_{\sma{\alpha,1}}}$, where $\alpha$ denotes the $\Z_N$ charge of $\ket{\phi_{\sma{\alpha,1}}^s}$. Since $[\calv_s,Q]=0$ for the quasi-adiabatic continuation of Eq.\ (\ref{qacmapping}), the $\Z_N$ charge of each state is preserved. It follows that
\begin{align} \label{calol}
&\bra{\phi^s_{\alpha,1}}\calo_l\ket{\phi^s_{\alpha+1\,\text{mod}\,N,1}}=1, \notag \\
\bra{\phi^s_{\alpha,1}&}\calo_r\ket{\phi^s_{\alpha+1\,\text{mod}\,N,1}}=\omega^{(N+1)/2+\alpha},
\end{align}
with the dressed operators
\begin{align} \label{definedressed}
\calo_l = \calv_s\gamma_{\sma{1}}\dg{\calv}_s \ins{and} \calo_r = \calv_s\pdg{\gamma}_{\sma{2L}}\dg{\calv}_s.
\end{align}
That $\calo_l$ is localized around site $1$ follows from $\calv_s$ being locality-preserving. Since both $\gamma_{\sma{1}}$ and $\calv_s$ are unitary,\ (\ref{calol}) also implies $\calo_l\ket{\phi^s_{\alpha+1,1}}=\ket{\phi^s_{\alpha,1}}$. \\

To gain some intuition about these edge modes, we construct $\calo_l$ for the simplest nontrivial example: a frustrated $\Z_2$ Kitaev chain, as modelled by 
\begin{align} \label{frustratedZ2model}
H_s = i\sum_{j=1}^{L-1}\gamma_{2j}\gamma_{2j+1}+is\sum_{j=1}^L\gamma_{2j-1}\gamma_{2j}.
\end{align}
Decomposing $H_s=H_0+sV$, we point out that $H_0$ is the frustration-free Kitaev model. The $V$ term tends to destabilize the topological phase, resulting in a monotonic decrease of the spectral gap $\Gamma^s$ above the two degenerate groundstates; it is known that $\Gamma^0=2$ in the frustration-free limit, and $\Gamma^1=0$, signalling a phase transition.\cite{fendley2012} The quasi-adiabatic continuation can thus be carried out for $s$ lying in the real interval $[0,\bar{s}]$, with $\bar{s}<1$ and $\Gamma^{\bar{s}}>0$. The quasi-adiabatic continuation operator has the form of a path-ordered evolution\cite{hastings2005}
\begin{align}  \label{qqq3}
\calv_s = T \text{exp}[i\int^s_0 \cald_{s'} ds']
\end{align}
in the `time' variable $s$, generated by the Hamiltonian
\begin{align} \label{qadiab3}
\cald_s = -i \int^{\infty}_{-\infty} dt\, F( \Gamma^s t)\, e^{iH_st}\,\big(\partial_sH_s\big)\,e^{-iH_st}.
\end{align}
We refer to $F(t)$ as a filter function, whose purpose is to cut off the time-evolution of $\partial_sH_s$ for large $|t|$. It is desirable that $F(t)$ has the fastest decay for large $|t|$, such that $\cald_s$ is maximally quasilocal;\cite{hastings2010,brayvi2011} this is elaborated in App.\ \ref{app:quasilocal}. In addition, $F(t)$ is imaginary, so that $\cald_s$ is Hermitian. Let us denote the eigenbasis of $H_s$ by $H_s \ket{j;s}=E_j^s\ket{j;s}$.  By choosing the Fourier transform of $F(t)$ as
\begin{align} \label{FTfilter}
\tilde{F}(\Omega) = \int^{\infty}_{-\infty} dt \,e^{i\Omega t}F(t) = -\frac{1}{\Omega} \;\;\text{for}\;\; |\Omega| \geq 1,
\end{align}
we obtain the matrix elements of $\cald_s$ as:
\begin{align}
i\bra{k;s}\cald_s\ket{j;s} = -\frac{\bra{k;s}\partial_sH_s\ket{j;s}}{E^s_k-E^s_j} = \bra{k;s}\partial_s\ket{j;s}
\end{align}
between states separated by the spectral gap $\Gamma_s$, i.e., for $|E^s_k-E^s_j|\geq \Gamma_s$.  This leads to $\partial_sP_s = i[\cald_s,P_s]$ and Eq.\ (\ref{qacmapping}). To first order in $s$, the edge mode has the form 
\begin{align} \label{edgemodefirstorder}
\calo_l = \calv_s \gamma_{\sma{1}} \dg{\calv}_s = \gamma_{\sma{1}}+ is[\cald_0,\gamma_{\sma{1}}] + \ldots. 
\end{align}
Employing $[H_0,\gamma_{\sma{1}}]=0$ and $[V,\gamma_{\sma{2j-1}}]=-2i\gamma_{\sma{j}}$, we are thus led to evaluate
\begin{align}
i[\cald_0,\gamma_{\sma{1}}] =-2i  \int^{\infty}_{-\infty} dt F(2t)\, e^{iH_0t}\, \gt\, e^{-iH_0t},
\end{align}
where we have identified $\Gamma^0=2$ as the spectral gap in the frustration-free Kitaev chain. Now apply that $\gt = (v_{\sma{+}}+v_{\sma{-}})/2$, where $v_{\sma{\pm}} = \gt \pm i \gth$ are constants under evolution by the dimerized $H_0$: $[H_0,v_{\sma{\pm}}]=\mp 2 v_{\sma{\pm}}$. We then find
\begin{align}
i[\cald_0,\gamma_{\sma{1}}] =&\; -i \int^{\infty}_{-\infty}  dt F(2t) \,\big(v_{\sma{+}}e^{-i2t}+v_{\sma{-}}e^{i2t}\big) =\gamma_{\sma{3}}. \notag
\end{align}
In the last step, we employed $\tilde{F}(\pm 1) = \mp 1$, as follows from Eq.\ (\ref{FTfilter}). We thus derive that the edge mode spreads as $\calo_l=\go + s\gth + \ldots$, and a simple computation shows $[H_s, \go +s\gth]=O(s^2)$. This result is suggestive of a two-fold degeneracy in the entire spectrum, as has been derived alternatively in Ref.\ \onlinecite{fendley2012} for higher orders in $s$. However, this property is not generic, as evidenced by our next illustration. In App.\ \ref{app:edgemode}, we work out the edge mode for the $\Z_3$ non-chiral parafermion, as modelled by $\hat{H}$ in Eq.\ (\ref{Z3example}) with $\hat{\phi}=0$. We identify the $J$-term as $H_0$, and $f$-term as the deformation $V$. The result is 
\begin{align} \label{edgemodeZ3}
\calo_l =&\; \pdg{\gamma}_{\sma{1}} + \tfrac{f}{3J}\big(\pdg{\gamma}_{\sma{3}} - \omega \dg{\gamma}_{\sma{2}}\dg{\gamma}_{\sma{3}}-\omega \dg{\gamma}_{\sma{1}}\dg{\gamma}_{\sma{3}} \big)\notag \\
&+ \tfrac{f}{3J}\, \omega^*\dg{\gamma}_{\sma{1}}\pdg{\gamma}_{\sma{2}}\pdg{\gamma}_{\sma{3}} + O\big( \tfrac{f^2}{J^2} \big),
\end{align}
which demonstrably does not commute with $H_s$. Instead, $\calo_l$ commutes with the projected Hamiltonian $P_sH_sP_s$ to superpolynomial accuracy, as follows from (i) $[\gamma_{\sma{1}},P_0]=0$ leading to $[\calo_l,P_s]=0$, and (ii) the groundstate degeneracy shown in Sec.\ \ref{sec:localindis}. In App.\ \ref{app:edgemode}, we have also derived the edge mode $\calo_l$ for the $\Z_3$ chiral parafermion ($0<\hat{\phi}<\pi/3$). In this parameter range, we find to first order in $s$ that $\calo_l$ may be deformed to commute with the full Hamiltonian $H_s$, at the cost of having a weaker decay away from the edge; this deformation involves changing only the filter function in the quasi-adiabatic Hamiltonian (\ref{qadiab3}). In this way, we recover  to first order the `zero edge mode' ($\Psi_{\sma{\text{left}}}$) that is alternatively derived in Eq. (32) of Ref.\ \onlinecite{fendley2012}; we do not know if this coincidence persists to higher orders. As described in App.\ \ref{app:caseb}, this deformation of the filter function becomes increasingly singular as $\hat{\phi} \rightarrow 0$, i.e., the deformed edge mode delocalizes in the non-chiral limit. In this sense, there is a trade-off between spatial localization and commutivity with the full Hamiltonian. \\

\begin{figure*}
\centering
\includegraphics[width=18 cm]{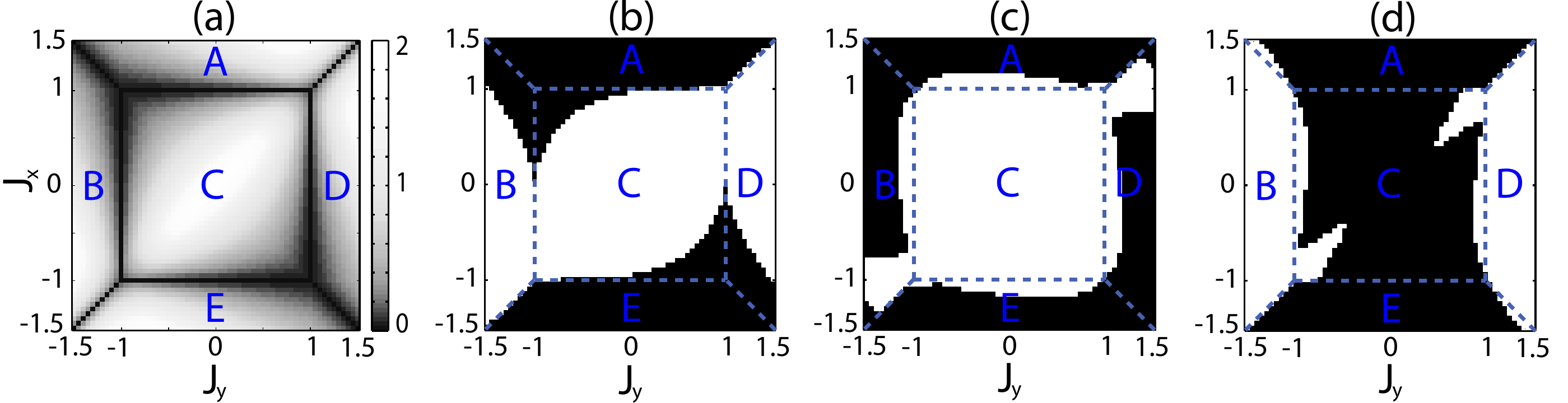}
\caption{ Characterizing the XYZ model on a chain of $L=22$ sites, with varying parameters $J_x$ and $J_y$. (a) The spectral gap above the two lowest states, for the open-chain Hamiltonian (\ref{xyzmajorana}). As delineated by gapless lines, the regions B, C and D correspond to topological phases, as explained in the main text. (b)-(d) evaluates various types of closed boundary conditions by the parity-switch criterion; parameter regions are colored white (resp.\ black) region where the groundstate parity switches (is invariant) upon twisting the boundary conditions. The blue dashed lines are phase transition lines on the \emph{open} chain. (b) Translational-invariant boundary conditions, as described in Eq.\ (\ref{xyzinteredgetrans}). (c) First-order topological boundary condition (TBC), i.e., Eq.\ (\ref{firstTBC}) with $\bar{\calq}^{\sma{(1)}}$ approximating Eq.\ (\ref{TOparameterXYZ}). This boundary condition is constructed for the topological region $C$ only. (d) First-order TBC with $\bar{\calq}^{\sma{(1)}}$ defined in App.\ \ref{app:regionBD}. This boundary condition is constructed for the topological regions $B$ and $D$ only.}\label{fig:xyzcomposite}
\end{figure*}

Let us present another application of these generalized edge modes. In the groundstate space, they generate a representation of the $\Z_N$ generator:
\begin{align} \label{Qfraction}
\calq = \omega^{(1-N)/2}\dg{\calo}_l\pdg{\calo}_r,
\end{align} 
which satisfies $\calq^N=I$, and $\calq\ket{\phi^s_{\alpha,1}} = \omega^{\alpha}\ket{\phi^s_{\alpha,1}}$. This may be compared with the representation\ (\ref{Zngenerator}) in the full Hilbert space, which operates on each site. The difference is that $\calq$ is a fractionalized representation, with support only near the chain edges.\cite{turner2011,bondesan2013,motruk2013} Together, $\calo_l$, $\calo_r$ and $\calq$ generate a non-commutative algebra in the groundstate space:
\begin{align} \label{algebra}
\calo_l &\calo_r = \omega \calo_r\calo_l, \;\; \calo_{l} \calq = {\omega} \calq \calo_{l},\notag \\
&\ins{and} \calo_{r} \calq = {\omega} \calq \calo_{r}.
\end{align}
For $N=2$, this is the familiar Clifford algebra, with $\calo_l$  acting as the Pauli matrix $\varsigma^x$ in the groundstate space, $\calo_r$ as $\varsigma^y$, and $\calq$ as $\varsigma^z$. Clearly, the charge of $\calo_l$ is a topological invariant\cite{turner2011,bondesan2013,motruk2013}, i.e., it does not change under quasi-adiabatic continuation. We say an operator $\calo$ has a charge $\alpha$ if $\dg{Q}\calo Q=\omega^{\alpha}\calo$. In the present context, $[Q,\calv_s]=0$ implies that $\calo_l$ has unit charge, independent of the deformation parameter $s$. A case in point is the $\Z_3$ edge mode\ (\ref{edgemodeZ3}), where each term in the expansion of $\calo_l$ has the same charge; the preceding discussion shows that this is true to all powers of $f/J$. \\

Finally, the existence of a fractionalized $\calq$ implies that the groundstate degeneracy is unstable to inter-edge coupling when we close the chain. It is worth refining our notion of locality on a closed chain. A case in point is $\calq$ in the frustration-free limit: 
\begin{align} \label{examplesupport}
\calq=\omega^{(1-N)/2}\dg{\gamma}_{\sma{1}}\pdg{\gamma}_{\sma{2L}} = \dg{\sigma}_1\sigma_L \prod_{j=1}^L\tau_j.
\end{align}
Since sites $1$ and $L$ are adjacent on a closed chain, $\calq$ is local in the parafermion representation, but nonlocal in the clock representation. In short, we call $\calq$ parafermion-local, but clock-nonlocal. Fractionalization implies the existence of a parafermion-local term that distinguishes the groundstates. That is, we may legitimately add to the open-chain Hamiltonian a term of the form: $\lambda \calq+h.c.$, which closes the chain and singles out a unique groundstate. Which groundstate is singled out depends on the phase factor $\lambda$, as we elaborate in Sec.\ \ref{app:detectTOonring}. We remark that this refined notion of locality is not necessary for an open-chain Hamiltonian, where charge-neutrality imposes that all parafermion-local terms are also clock-local; this is proven in App.\ \ref{app:quasilocal}. Finally, we point out that the local-indistinguishability condition\ (\ref{spo}) on an open chain applies only for symmetric, \emph{clock}-local probes, as may be verified in the proof of Sec.\ \ref{sec:localindis}; this allows for a clock-nonlocal but parafermion-local term, e.g. $\calq$, to close the chain and break the groundstate degeneracy.

\section{Topological order on a closed chain} \label{app:detectTOonring}

The topological degeneracy that we attribute to edge modes is generically lost on a closed chain.\cite{motruk2013,bondesan2013} One might therefore ask: what inter-edge coupling guarantees that the nondegenerate, closed-chain groundstate ($|\Phi\rangle$) is topological, i.e., that $|\Phi\rangle$ is continuously connected to the topological groundstate manifold of the open chain? Does there exist a coupling that uniquely selects a closed-chain groundstate to be \emph{exactly} one of the open-chain groundstates? The goal of Sec.\ \ref{twist:qc} is to prove the existence of this inter-edge coupling for Majorana chains, and to describe how the coupling is constructed. In Sec.\ \ref{sec:transinvBC}, we perform a critical comparison with the conventional boundary condition that is commonly employed. We then extend our discussion to $\Z_N$ parafermions in Sec.\ \ref{sec:prooftwist}.

\subsection{Topological boundary conditions} \label{twist:qc}

The simplest illustration lies in the frustration-free Majorana model, whose Hamiltonian on an open chain is 
\begin{align} \label{H0maj}
H_0 = i\sum_{j=1}^{L-1}\pdg{\gamma}_{\sma{2j}}\pdg{\gamma}_{\sma{2j+1}}.
\end{align}
As noted in the previous Section, the groundstate manifold is topologically ordered, and comprises a state $|\phi_{\sma{\alpha,1}}\rangle$ in each charge sector $\alpha \in \Z_2$. We would like to construct a closed-chain Hamiltonian ($\calh^{\sma{(\alpha)}}_{\sma{0}}$) that is {exactly} minimized by the open-chain groundstate $|\phi_{\sma{\alpha,1}}\rangle$. This is accomplished by identifying the low-energy degree of freedom on the open chain, which in this example is the complex fermion $\tilde{c} =(\gamma_{\sma{1}}-i\gamma_{\sma{2L}})/2$. It is simple to verify that $|\phi_{\sma{0,1}}\rangle$ and $|\phi_{\sma{1,1}}\rangle$ differ in their occupation of $\tilde{c}$. Therefore, we are led to $\calh^{\sma{(\alpha)}}_{\sma{0}} =H_{\sma{0}} + \delta H^{\sma{(\alpha)}}_{\sma{0}}$, with the inter-edge coupling
\begin{align} \label{twistmajorana}
\delta H_0^{{(\alpha)}}=i(-1)^{\alpha}\pdg{\gamma}_{\sma{1}}\pdg{\gamma}_{\sma{2L}} = (-1)^{\alpha}(1-2\dg{\tilde{c}}\tilde{c}).
\end{align}
The phase factor $(-1)^{\sma{\alpha}}$ may be interpreted as flux insertion, and determines the occupation of the fermion ($\tilde{c}$) in the nondegenerate groundstate. Alternatively stated, we obtain two Hamiltonians by imposing periodic and antiperiodic boundary conditions for the complex fermion $c_{\sma{j}} =(\gamma_{\sma{2j-1}}-i\gamma_{\sma{2j}})/2$, as schematically illustrated in Fig.\ \ref{fig:moebius}. \\

Let us generalize our discussion to frustrated models that remain topologically ordered. For illustration, we may deform the open-chain $H_0$ as:
\begin{align} \label{xyzmajorana}
H_{\sma{XYZ}} =&\; \sum_{j=1}^{L-1} \big(\, J_x\pdg{\gamma}_{\sma{2j-1}}\pdg{\gamma}_{\sma{2j}}\pdg{\gamma}_{\sma{2j+1}}\pdg{\gamma}_{\sma{2j+2}} \notag \\
-iJ_y&\pdg{\gamma}_{\sma{2j-1}}\pdg{\gamma}_{\sma{2j+2}} +i\pdg{\gamma}_{\sma{2j}}\pdg{\gamma}_{\sma{2j+1}}\,\big).
\end{align}
We refer to this as the XYZ model, as it can be expressed as a spin-half model through the Jordan-Wigner transformation\ (\ref{pfclocktransformation}); cf. App.\ \ref{app:duality}. As shown in Fig.\ \ref{fig:xyzcomposite}(a) for the range $|J_x|,|J_y|<1$ (square labeled C), the spectral gap persists above the two lowest-lying states, which we denote by $|\phi^{\sma{J_x,J_y}}_{\sma{\alpha,1}}\rangle$ in the charge sector $\alpha$. The low-energy degree of freedom on the open chain is directly generalized as $\tilde{c}= (\calo_{\sma{l}}-i\calo_{\sma{r}})/2$, where we have dressed the frustration-free edge modes as $\gamma_{\sma{1}} \rightarrow \calo_{\sma{l}}(J_x,J_y)$ and $\gamma_{\sma{2L}} \rightarrow \calo_{\sma{r}}(J_x,J_y)$, following Eq.\ (\ref{definedressed}). Such a dressing exists for any deformation within the square C, where a gapped interpolation exists to the frustration-free limit. In the presence of interactions, $\tilde{c}$ represents a many-body excitation. We then close the chain with the inter-edge coupling
\begin{align} \label{effectiveHamtwochains}
\delta H_{\sma{XYZ}}^{{(\alpha)}}=(-1)^{\alpha+1}\calq = (-1)^{\alpha}(1-2\dg{\tilde{c}}\tilde{c}),
\end{align}
where $\calq=-i{\calo}_{\sma{l}}\pdg{\calo}_{\sma{r}}$ is derived for the XYZ model to be
\begin{align} \label{TOparameterXYZ}
&\calq = -i\go \gtL + iJ_y(\gfi \gtL + \go \gtLmfo) \notag \\
&\;\; -J_x (\gt \gth \gfi \gtL + \go \gtLmfo \gtLmt \gtLmo) + \ldots,
\end{align}
to first order in the deformation parameters; cf. App.\ \ref{app:xyztwist}. Apparently, $\calq$ cannot be derived by translating the bulk terms of Eq.\ (\ref{xyzmajorana}) to the edge. Since $\calq$ represents the fermion parity in the open-chain groundstate space (cf. Sec.\ \ref{sec:zeromodes}),
\begin{align}
\delta H_{\sma{XYZ}}^{{(\alpha)}}\ket{\phi^{\sma{J_x,J_y}}_{\beta,1}}= (-1)^{\alpha+\beta+1}\ket{\phi^{\sma{J_x,J_y}}_{\beta,1}},
\end{align}
for $|J_x|,|J_y|<1$. This implies  that $\ket{\phi^{\sma{J_x,J_y}}_{\sma{\alpha,1}}}$ separately minimizes the open-chain Hamiltonian ($H_{\sma{XYZ}}$) and the inter-edge coupling ($\delta H_{\sma{XYZ}}^{\sma{(\alpha)}}$), i.e., $\ket{\phi^{\sma{J_x,J_y}}_{\sma{\alpha,1}}}$ must be the nondegenerate groundstate of the closed-chain Hamiltonian ($H_{\sma{XYZ}}+\delta H_{\sma{XYZ}}^{\sma{(\alpha)}}$). A corollary is that the closed-chain groundstate switches parity on twisting the boundary conditions, while the groundstate energy is preserved. In short, we call this permutation-by-twisting. \\

While we have focused on one model for specificity, permutation-by-twisting applies more generally to any closed-chain Hamiltonian that may be decomposed as 
\begin{align} \label{TBC}
\tilde{\calh}^{{(\alpha)}} = H_{\sma{open}} + (-1)^{\alpha+1}\calq. 
\end{align}
Here, $H_{\sma{open}}$ is any open-chain Hamiltonian which is quasi-adiabatically connected to $H_0$, and $\calq=-i{\calo}_{\sma{l}}\pdg{\calo}_{\sma{r}}$ is the fractionalized representation of the fermion parity in the groundstate space of $H_{\sma{open}}$. We have thus established a bulk-edge correspondence: between the existence of topological edge modes on an open chain, to the property of permutation-by-twisting on a closed chain. We refer to $\calq$ as a topological order parameter, in analogy with the clock-local order parameter ($\hat{O}$) of traditional broken-symmetry systems. There, the perturbation $\lambda \hat{O}+h.c.$ picks out a broken-symmetry groundstate which depends on $\lambda$. The difference is that $\hat{O}$ breaks the symmetry but preserves clock-locality, while $\calq$ preserves the symmetry and parafermion-locality, but breaks clock-locality.\\

In practice, the exact form of $\calq$ is not easily found. However, the first-order truncation ($\bar{\calq}^{\sma{(1)}}$) of $\calq$ is analytically tractable, e.g., Eq.\ (\ref{TOparameterXYZ}) without the dots is an approximation to $\calq$ to first order in the deformation parameters ($J_x,J_y$). We thus describe the closed-chain Hamiltonian
\begin{align} \label{firstTBC}
\calhxyz^{(\alpha)} = \hxyz - (-1)^{\alpha}\bar{\calq}^{\sma{(1)}}
\end{align}
as having a first-order topological boundary condition (TBC), which we proceed to evaluate. Here and in future sections, we use  the parity switch as the simplest diagnostic to evaluate how `topological' a closed-chain Hamiltonian is. Ideally, the nondegenerate groundstate parity switches upon twisting the boundary condition, for all parameters where the open-chain Hamiltonian is topologically-ordered. Indeed, we show in Fig.\ \ref{fig:xyzcomposite}(c) that the parity switch occurs (whitened regions) in nearly the entire square $C$. We remark that the boundary condition\ (\ref{firstTBC}) is specifically constructed for the topological region $C$. As we will shortly clarify, regions $B$ and $D$ in Fig.\ \ref{fig:xyzcomposite} are also topological, but they require a different set of TBC's.\\

\begin{figure}[h]
\centering
\includegraphics[width=8.3 cm]{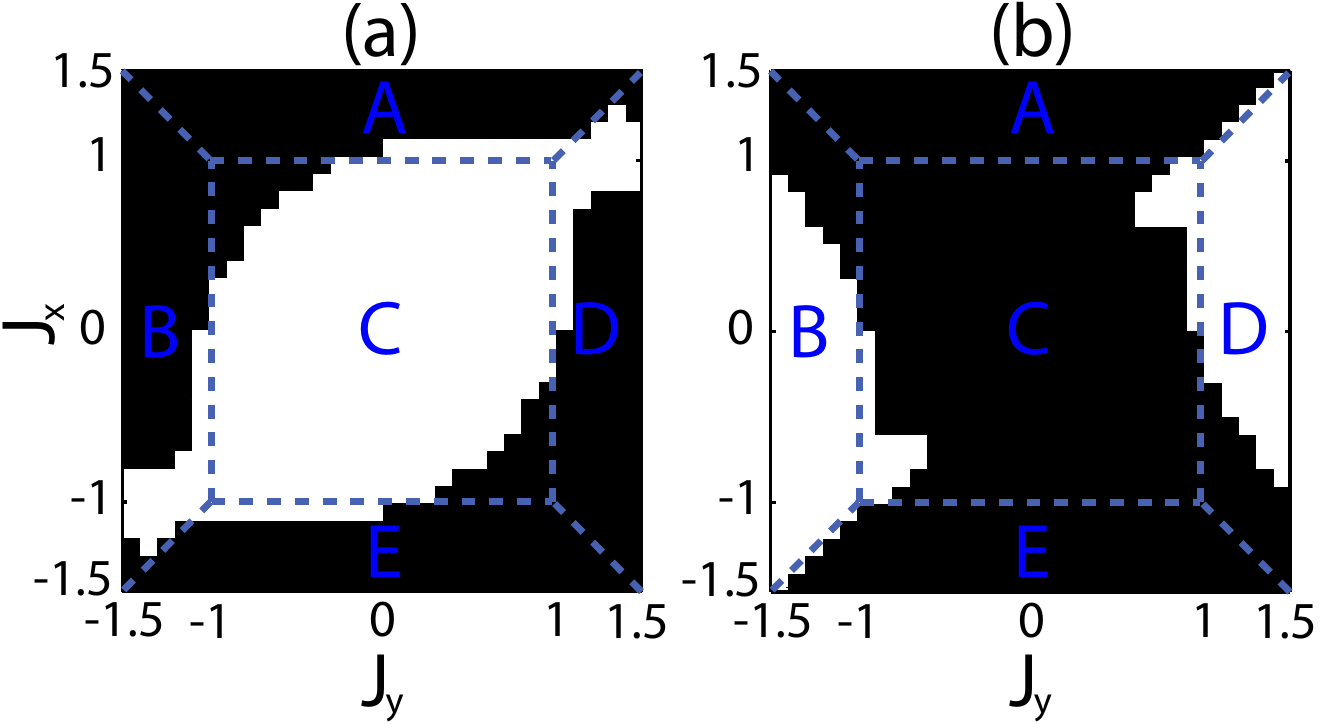}
\caption{ Evaluation of the zeroth-order topological boundary conditions, for the XYZ model on a closed chain of $L=22$ sites. (a) We employ the inter-edge coupling $(-1)^{\sma{\alpha}}i\go \gtL$. (b) Here, we employ  $(-1)^{\sma{\alpha}}J_y\,i\gt \gtLmo$. White (resp.\ black) region: the groundstate parity switches (is invariant) upon twisting the boundary conditions.}\label{fig:xyz_zeroth}
\end{figure}

To critically evaluate the effect of the first-order dressing, we also consider a boundary condition that is zeroth-order in $J_x$ and $J_y$, i.e., we apply the inter-edge coupling\ (\ref{TBC}) with $\calq$ replaced by $\bar{\calq}^{\sma{(0)}}=-i\go \gtL$. For this zeroth-order TBC, we compute the corresponding parity-switch diagram in Fig.\ \ref{fig:xyz_zeroth}(a). We find that the parity switch remains robust along $-1<J_x=J_y <1$, where the open-chain groundstates are completely classical; the groundstates of the ferromagnetic XXZ model are well-known\cite{Mikeska} to be fully-polarized. In this classical limit, one may verify that $\bar{\calq}^{\sma{(0)}}$ and $\calq$ have identical matrix elements within the groundstate space. This coincidence is lost away from the XXZ line, where quantum fluctuations play an important role. Where fluctuations are strongest (in the orthogonal direction: ${-1}<J_x={-J_y} <1$), the parity switch does not always occur -- the first-order TBC significantly outperforms its zeroth-order counterpart.

To conclude our discussion of the XYZ model, we remark that regions $B$ and $D$ (see Fig.\ \ref{fig:xyzcomposite}) \emph{also} correspond to a topological phase, by their quasi-adiabatic connection to a \emph{different} dimer Hamiltonian: 
\begin{align}
H_0'=  -iJ_y\sum_{j=1}^{L-1}\pdg{\gamma}_{\sma{2j-1}}\pdg{\gamma}_{\sma{2j+2}},
\end{align}
which has edge modes: $\gt$ and $\gtLmo$. Consequently, the TBC in regions $B$ and $D$ differs from Eq.\ (\ref{TOparameterXYZ}), as we show in App.\ \ref{app:regionBD}. By the parity-switch criterion, the first-order TBC (Fig.\ \ref{fig:xyzcomposite}(d)) is shown to be more effective than the zeroth-order TBC's (Fig.\ \ref{fig:xyz_zeroth}(b)). 

\subsection{Comparison with the translational-invariant boundary condition} \label{sec:transinvBC}

Having introduced a novel type of boundary condition, we would like to make a critical comparison with the conventional method to close chains. Assuming that the open-chain Hamiltonian ($H_{\sma{open}}$) is translational-invariant up to the edge, it is common practice\cite{fidkowski2011b,zaletel2014} to translate the bulk terms of $H_{\sma{open}}$ to the edge, and then add phase factors for different flux insertions. For the XYZ model, this procedure produces the closed-chain Hamiltonian:
\begin{align} \label{xyzinteredgetrans}
&\calk^{{(\alpha)}}_{\sma{XYZ}} = \hxyz + J_x\pdg{\gamma}_{\sma{1}}\pdg{\gamma}_{\sma{2}}\pdg{\gamma}_{\sma{2L-1}}\pdg{\gamma}_{\sma{2L}} \notag \\
&-i (-1)^{\alpha}J_y \pdg{\gamma}_{\sma{2}}\pdg{\gamma}_{\sma{2L-1}} + i (-1)^{\alpha} \pdg{\gamma}_{\sma{1}}\pdg{\gamma}_{\sma{2L}},
\end{align}
as we systematically derive in App.\ \ref{app:twistmethod}; there, we also generalize this procedure for $\Z_N$ parafermions. By construction, $\calk^{\sma{(\alpha)}}$ is translational-invariant modulo a $\pm 1$ phase factor. A stronger statement is that $\calk^{\sma{(0)}}$ (resp.\ $\calk^{\sma{(1)}}$) is completely translational-invariant in the even (resp.\ odd) charge sector, as we demonstrate in App.\ \ref{app:duality}. Therefore, we describe $\kal$ as having a translational-invariant boundary condition (TIBC). In comparison, the TBC generically breaks translational invariance, as exemplified in Eq.\ (\ref{TOparameterXYZ}). \\

It is commonly believed that if $H_{\sma{open}}$ is topologically-ordered, then the closed-chain $\calk^{\sma{(\alpha)}}$ must also be topological, implying that the groundstate of $\calk^{\sma{(\alpha)}}$ is especially sensitive to its boundary conditions.\cite{fidkowski2011b,zaletel2014} We numerically test this expectation for the XYZ model on a finite-size chain -- see Fig.\ \ref{fig:xyzcomposite}(b). The groundstate parity is invariant where quantum fluctuations are dominant, as we have also seen qualitatively for the zeroth-order TBC (Fig.\ \ref{fig:xyz_zeroth}(a)); both these boundary conditions are outperformed by the first-order TBC (see Fig.\ \ref{fig:xyzcomposite}(c)). A more thorough finite-size analysis in App.\ \ref{app:duality} suggests these conclusions are robust in the thermodynamic limit.

\subsection{Closing the chain for general $\Z_{N>2}$ parafermions} \label{sec:prooftwist} 

For $\Z_{\sma{N>2}}$ parafermions on an open chain, the low-energy degree of freedom is no longer a complex fermion $\tilde{c}$. Nevertheless, the $N$ eigenvalues of $\calq$ correspond to distinct states in the groundstate manifold, and the generalized edge modes ($\calo_l$, $\calo_r$) induce many-body excitations within this manifold, as per the algebra in Eq.\ (\ref{algebra}). To select a topological groundstate on a closed chain, we propose the inter-edge coupling:
\begin{align} \label{ZNinteredge}
\delta H^{(\alpha)} = - \frac1{N}\sum_{\beta=0}^{N-1} \big(\,\omega^{\alpha}\calq\,\big)^{\beta},
\end{align}
which projects to the nondegenerate groundstate with charge $N-\alpha$, for $\alpha \in \Z_N$. Twisting the Hamiltonian $(\delta H^{\sma{(\alpha)}} \rightarrow \delta H^{\sma{(\alpha+1)}})$ cyclically permutes the $\Z_N$ charge of the nondegenerate groundstate, while preserving the groundstate energy. \\

To see this, let us label the open-chain groundstates as $|\phi^s_{\sma{\alpha,1}}\rangle$ in the charge sector $\alpha$; recall that $s$ parametrizes the deformation of the open-chain Hamiltonian: $H_s=H_0+sV$. By construction, $\delta H^{\sma{(\alpha)}}$ singles out $|\phi^s_{\sma{N-\alpha,1}}\rangle$ as having the lowest eigenvalue: $\delta H^{\sma{(\alpha)}}  |\phi^s_{\sma{N-\alpha,1}}\rangle = -|\phi^s_{\sma{N-\alpha,1}}\rangle$. $|\phi^s_{\sma{N-\alpha,1}}\rangle$ has to be the nondegenerate groundstate of $H_{\sma{op}}+\delta H^{\sma{(\alpha)}}$. This follows from $\calo_l$, $\calo_r$ and $\calq$ being unitary, which implies the operator norm $||\delta H^{\sma{(\alpha)}}||_{\sma{\text{op}}} \leq 1$ by the triangle inequality. Thus, $|\phi^s_{\sma{N-\alpha,1}}\rangle$ separately minimizes $H_{\sma{op}}$ and $\delta H^{\sma{(\alpha)}}$.\\

\begin{figure}[h]
\centering
\includegraphics[width=8.3 cm]{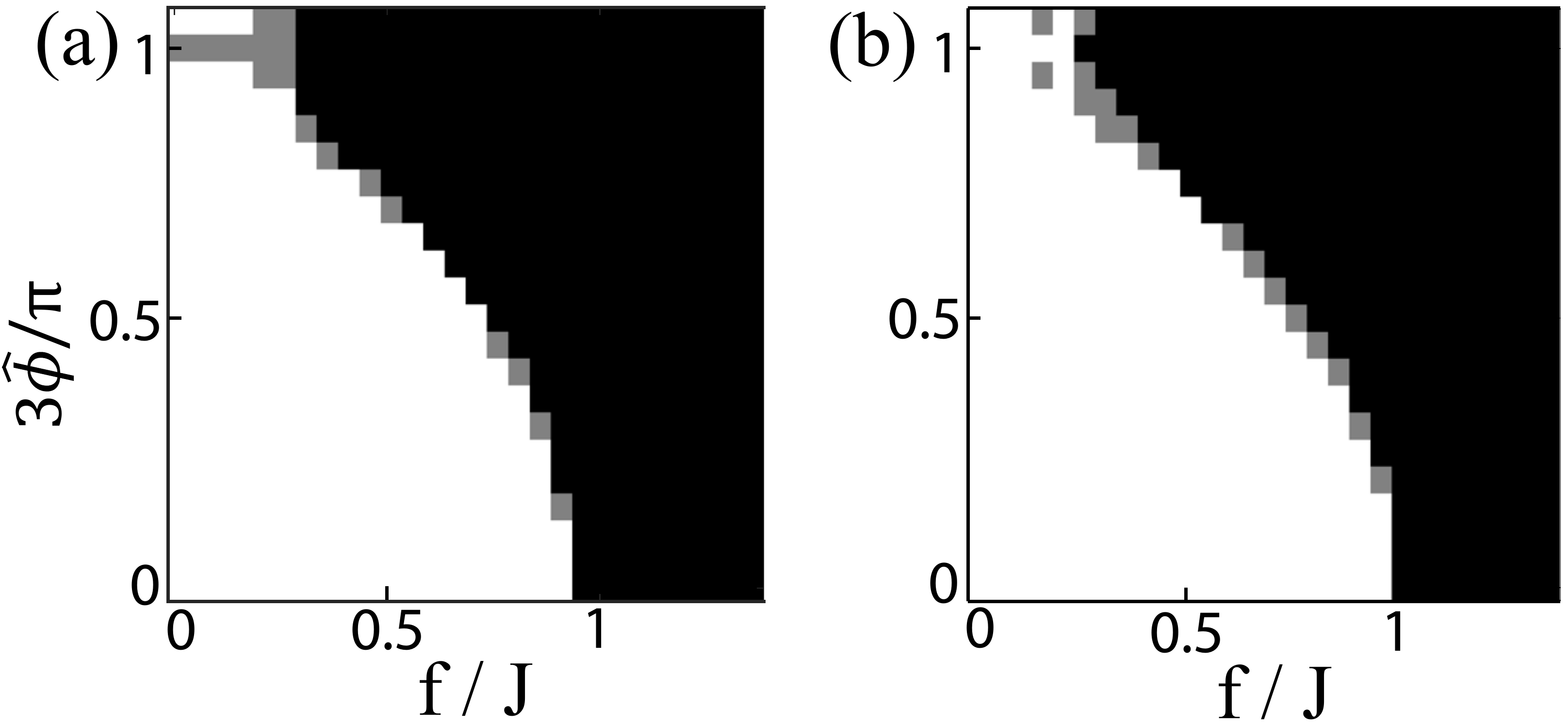}
\caption{ Simulation of the $\Z_3$ model (\ref{Z3example}) on a $14$-site chain. (a) The chain is closed with an inter-edge coupling $\delta H^{\sma{(\alpha)}} = -\omega^{\sma{\alpha-1}}\dg{\go}\gtL/3 +h.c.$, which we call zeroth-order TBC. The figure represents the nondegenerate-groundstate charge $Q_0^{\sma{(\alpha)}}$, as we twist the closed-ring Hamiltonian through all $\Z_3$ values of the twist parameter $\alpha$. White region: $Q_0$ is maximally permuted, i.e., $\{Q_0^{\sma{(\alpha)}}\}$ takes on three values, indicating a topological phase. Gray: two. Black: one. (b) For comparison, we have also simulated the translational-invariant boundary condition, which involves the coupling $\delta H^{\sma{(\alpha)}} = -J\omega^{\sma{\alpha-1}}e^{i\hat{\phi}}\dg{\go}\gtL +h.c.$.  }\label{fig:Z3twist}
\end{figure}

For illustration, we return to the $\Z_3$, open-chain model (\ref{Z3example}), and now add an inter-edge coupling in the form of Eq.\ (\ref{ZNinteredge}). In the frustration-free limit ($f=0$), the edge modes are $\go$ and $\gtL$, and therefore the topological order parameter equals $\omega^*\dg{\go}\gtL$. In our numerical simulation, we ignored the dressing of $\calq$ for nonzero $f$, and used its frustration-free form throughout; once again, we refer to this as zeroth-order TBC. Despite this approximation, the zeroth-order TBC correctly identifies the topological region in the range of Hamiltonians that we explored; see Fig.\ \ref{fig:Z3twist}(a), where the topological region (colored white) indicates a $\Z_3$ permutation of the non-degenerate groundstate charge, as we twist the boundary conditions. To support our claim that this region is topological, we note its close overlap with the white region of Fig.\ \ref{fig:Z3}, where the groundstate is three-fold degenerate on an open chain.

\section{Generalization to phases with symmetry-breaking} \label{sec:pfsymmbreak}

A new type of phase can occur for $\Z_4$ parafermions, where the groundstate $|\theta\rangle$ has a reduced $\Z_2$ symmetry. This means that $|\theta\rangle$ no longer has a conserved charge under $Q$, but it does so with $Q^2$, i.e., it is a condensate\cite{motruk2013} of parafermion pairs. A general classification of $\Z_N$ parafermions allows for these coherent states, sometimes in conjunction with topological order. All distinct phases are uniquely labelled by $n$, which is a divisor of $N$.\cite{bondesan2013,motruk2013} The $n=1$ phase is purely topological, and has been characterized in the previous Sections; we now turn our attention to $1<n<N$. It is useful to define $m=N/n$ and the greatest common divisor of $m$ and $n$: $g=\text{gcd}(m,n)$. If $g>1$, there is symmetry-breaking with a parafermionic order parameter; if $m>g$, the phase is topological ordered; if $m>g>1$, symmetry-breaking coexists with topological order. A major advance of Ref.\ \onlinecite{bondesan2013} is the construction of frustration-free models which realizes each of these phases on an open chain:
\begin{align} \label{pfHam}
\hnn &\;= -\frac{1}{m} \sum_{j=1}^{L-1} \sum_{\beta=0}^{m-1}(\omega^{(N-1)/2} \dg{\gamma}_{\sma{2j+1}}\pdg{\gamma}_{\sma{2j}})^{\beta n} \notag \\
&- \frac{1}{n}\sum_{j=1}^{L}\sum_{\alpha=0}^{n-1}(\omega^{(1-N)/2}\dg{\gamma}_{\sma{2j-1}} \pdg{\gamma}_{\sma{2j}})^{\alpha m}.
\end{align}
For example, $H_{\sma{(4,2)}}$ realizes the above-mentioned condensate of parafermion pairs. $\hnn$ is a sum of mutually-commuting projections, and generalizes the dimer model\ (\ref{pfHamn1}). Many groundstate properties are now known in the frustration-free limit,\cite{bondesan2013} including the form of the parafermionic order parameters and the topological edge modes; these properties are reviewed in Sec.\ \ref{sec:reviewpfgeneraln}. The aim of Sec.\ \ref{sec:condensatestabilityopen} is to demonstrate that the phases remain stable even if frustrated on an open chain. Sec.\ \ref{sec:condensatestabilityclosed} extends the stability analysis to a closed chain. 

\subsection{Review of general parafermionic phases} \label{sec:reviewpfgeneraln}

We begin by describing the groundstate properties of $\hnn$, of which many are known from Ref.\ \onlinecite{bondesan2013}, then extend them to more general Hamiltonians by quasi-adiabatic continuation. It is instructive to interpret $\hnn$ in the clock representation, where \emph{all} nontrivial phases arise from traditional symmetry-breaking with a clock-local order parameter. For illustration, we consider the clock-analog of the parafermion-pair condensate. One possible groundstate of $H_{\sma{(4,2)}}$ is fully-polarized as $\bigotimes_{\sma{j=1}}^{\sma{L}}(|0\rangle_{\sma{j}}+|2\rangle_{\sma{j}})$. This state breaks the $Z_4$ symmetry under $|\alpha\rangle_{\sma{j}} \rightarrow |\alpha +1\; \text{mod}\; 4\rangle_{\sma{j}}$, but retains a reduced symmetry under $|\alpha\rangle_{\sma{j}} \rightarrow |\alpha +2\; \text{mod} \; 4\rangle_{\sma{j}}$. More generally,
\begin{align} \label{HNnclock}
\hnn=-\frac{1}{m}\sum_{j=1}^{L-1}\sum_{\beta=0}^{m-1}(\pdg{\sigma}_j\dg{\sigma}_{j+1})^{\beta n} -  \frac{1}{n}\sum_{j=1}^{L}\sum_{\alpha=0}^{n-1}\tau_j^{\alpha m}
\end{align}
has $m$ classical groundstates $|\psi_{\sma{\alpha,n}}\rangle$, which are fully polarized in the quantum numbers of $\sigma_j^n$: 
\begin{align} \label{psin}
&\ket{\psi_{\alpha,n}} = \bigotimes_{j=1}^L\frac{1}{\sqrt{n}}\sum_{\beta=0}^{n-1}\ket{\alpha+m\beta}_j, 
\end{align}
where $\alpha  \in \Z_m$, and $\sigma_{\sma{j}}\ket{\alpha}_{\sma{j}}=\omega^{\alpha}\ket{\alpha}_{\sma{j}}$. The $\Z_N$ generator\ (\ref{Zngenerator}) acts in this basis as $Q|\psi_{\sma{\alpha,n}}\rangle = |\psi_{\sma{\alpha+1 \,\text{mod}\,m,n}}\rangle$, as follows from\ (\ref{basisonsite}). In the charge eigenbasis, 
\begin{align} \label{chargeeigenbasis}
\ket{\phi_{\alpha,n}} = \frac{1}{\sqrt{m}}\sum_{\beta=0}^{m-1}{\omega}^{-n \alpha \beta} \ket{\psi_{\beta,n}} 
\end{align}
with $\alpha \in \Z_m$ and $Q|\phi_{\sma{\alpha,n}}\rangle = \omega^{n\alpha}|\phi_{\sma{\alpha,n}}\rangle$. Since $|\psi_{\sma{\alpha,n}}\rangle$ are fully-polarized in the clock representation, clock-local operators have difficulty transforming one $|\psi\rangle$ into another; this transformation would involve turning the clocks on all sites. This observation leads to the local-indistinguishability of the groundstate space $\pnn = \sum_{\alpha=0}^{m-1}|\phi_{\sma{\alpha,n}}\rangle \langle \phi_{\sma{\alpha,n}}|$, i.e., $\pnn$ satisfies the same condition as $\pno$ in Eq.\ (\ref{spo}),  when probed by any symmetric, clock-local operator ($\calo$). To show this, we first suppose the probe $\calo$ has a finite range. We would like to show that the matrix $\calm^{\sma{(n)}}_{\sma{\ab}}=\langle \phi_{\sma{\alpha,n}}|\,\calo\,|\phi_{\sma{\beta,n}}\rangle$ is proportional to the identity. The off-diagonal elements vanish because $[\calo,Q]=0$. The difference in diagonal elements, $\calm^{\sma{(n)}}_{\sma{\alpha \alpha}}- \calm^{\sma{(n)}}_{\sma{\beta \beta}}$, equals a sum of terms proportional to $\langle \psi_{\sma{\mu,n}}| \,\calo\, |\psi_{\sma{\nu,n}}\rangle$, with $\mu \neq \nu$. This quantity vanishes due to the above-mentioned observation. If $\calo$ is not finite-ranged but decays superpolynomially, $\calm^{\sma{(n)}}_{\sma{\alpha \alpha}}- \calm^{\sma{(n)}}_{\sma{\beta \beta}}$ vanishes up to finite-size corrections that we bound in App.\ \ref{app:generalprooflocalindist}. An instructive basis for $\pnn$ is 
\begin{align} \label{instructivebasis}
\ket{\theta_{\beta,\delta}} =   \frac{1}{\sqrt{g}}\sum_{\alpha=0}^{g-1}e^{-i2\pi\alpha \beta /g} \ket{\phi_{\alpha m/g+\delta,n}},
\end{align}
where $\alpha m/g+\delta \in \Z_m$, $\beta \in \Z_g$ may be interpreted as a broken-symmetry index, and $\delta \in \Z_{m/g}$ as a topological index; recall $g=\text{gcd}(m,n)$. On application of a symmetry-reducing on-site `field': $\lambda\gamma_{\sma{j}}^{\sma{N/g}}+h.c.$, the $m$-multiplet splits into $g$ number of $(m/g)$-multiplets, as determined by 
\begin{align} \label{pfreduce}
&\gamma_{\sma{2j-1}}^{\sma{N/g}}\ket{\theta_{\beta,\delta}} = e^{-i 2\pi \beta /g}\ket{\theta_{\beta,\delta}}, \;\;\text{and} \notag \\
\gamma_{\sma{2j}}^{\sma{N/g}}&\ket{\theta_{\beta,\delta}} = \omega^{N^2(g-1)/2g^2} e^{-i 2\pi \beta /g}\ket{\theta_{\beta,\delta}}.
\end{align}
For any subcell $j$, $\gamma_{\sma{j}}^{\sma{N/g}}$ are parafermionic order parameters which reduce the $\Z_N$ symmetry of $\hnn$ to $\Z_{N/g}$, i.e., each $(m/g)$-multiplet has a residual symmetry generated by $Q^g$. If $m>g$, there exists a remnant $(m/g)$-fold degeneracy which originates from topological edge modes. These localized operators ($\gamma_{\sma{1}}^n$ and ${\gamma}_{\sma{2L}}^n$) commute with $\hnn$ and permute the groundstates as $|\theta_{\sma{\beta,\delta}}\rangle \rightarrow |\theta_{\sma{\beta,\delta-1 \,\text{mod}\, m/g}}\rangle$. In App.\ \ref{app:modelcoexist}, we exemplify this discussion with the model $\het$, which has coexisting topological and symmetry-breaking orders. \\

The exact form of $|\theta \rangle$ allows us to expand upon their interpretation\cite{motruk2013} as coherent states. Eq.\ (\ref{pfreduce}) implies that $|\theta\rangle$ is a condensate of the operator $\varphi_{\sma{j}} =\gamma_{\sma{j}}^{\sma{N/g}}$. This condensate is not of Bose-Einstein type, since
\begin{align}
[\gamma_{\sma{i}}^{\sma{pN/g}},\gamma_{\sma{j}}^{\sma{qN/g}}]=0 \ins{for} p,q \in \Z,\;\forall\; i,j,
\end{align}
implying that $[\varphi_{\sma{i}},\dg{\varphi}_{\sma{j}}]=0 \neq \delta_{ij}$. Nevertheless, off-diagonal long-range order manifests as $\langle \theta_{\sma{\beta,\delta}}|\, \dg{\varphi}_{\sma{2i-1}} \varphi_{\sma{2j-1}} \, |\theta_{\sma{\beta,\delta}}\rangle = 1$, independent of $|i-j|$. Note that $[\varphi_{\sma{i}},{\varphi}_{\sma{j}}]=[\dg{\varphi}_{\sma{i}},\dg{\varphi}_{\sma{j}}]=0$, as would a bosonic operator. The present situation is reminiscent of long-range order in the BCS wavefunction, where the order parameter is almost bosonic. There are differences: (a) the groundstate expectation $\langle \theta_{\sma{\beta,\delta}}| \,\varphi_{\sma{j}} \, |\theta_{\sma{\beta,\delta}} \rangle$ takes on $g$ discrete values (cf. Eq.\ (\ref{pfreduce})), in contradistinction with conventional BCS wavefunctions that have a $U(1)$ degree of freedom. (b) The order parameter $\varphi$ manifests an attraction between $N/g$ parafermions $(g>1)$, thus generalizing Cooper pairs to `Cooper multiplets' if $N/g>2$.

\subsection{Robustness of general parafermionic phases on an open chain} \label{sec:condensatestabilityopen}

Let us address the stability of these phases as we symmetrically deform $\hnn$ to a new Hamiltonian $H_s$. As long as the deformation preserves the gap above the lowest $m$ states, there exists a quasi-adiabatic continuation ($\calv_s$) which maps their respective groundstate spaces as $\pnn$ to $P_s$. Thus if $\pnn$ is indistinguishable to clock-local probes (as shown in Sec.\ \ref{sec:reviewpfgeneraln}), then this property robustly carries forward to $P_s$, following a simple generalization of Sec.\ \ref{sec:localindis}. Moreover, in close analogy with Sec.\ \ref{sec:zeromodes}, we find for $g>1$: generalized order parameters $\caly_j=\calv_s\varphi_{\sma{j}}\dg{\calv}_s$, which are dressed versions of $\varphi_{\sma{j}}=\gamma_{\sma{j}}^{\sma{N/g}}$. Dressing $\caly_j$ may be interpreted as spreading the Cooper-multiplet wavefunction, in contrast with the tightly-bound $\varphi_{\sma{j}}$. By construction, $\caly_j$ permutes the groundstates of $H_s$, in the same manner that $\varphi_{\sma{j}}$ would for $\hnn$. While $[\varphi_{\sma{j}},\hnn]=0$, $\caly_j$ commutes only with the groundstate-projected $H_s$. \\

If the deformed Hamiltonian $H_s$ is topologically-ordered, there exist generalized edge modes  $\{\calo_{l,n},\calo_{r,n}\}$ which permute the groundstates, and are related by quasi-adiabatic continuation to  $\{\gamma_{\sma{1}}^{n},{\gamma}_{\sma{2L}}^{n}\}$; here, the subscript $l$ (resp.\ $r$) indicates that the operator is localized on the left (resp.\ right) edge. These edge modes generate a fractionalized representation of the $\Z_{N/g}$ generator:
\begin{align}
\calq_g = \omega^{xn(xn-N)/2} (\calo_{l,n})^{-x}(\calo_{r,n})^{x},
\end{align}
with integer $x$ uniquely satisfying $xn=g$ mod $m$. This operator acts like $Q^g$ in the groundstate space of $\hnn$, i.e.,  in the basis\ (\ref{instructivebasis}),
\begin{align} \label{actionQg}
\calq_g\ket{\theta_{\beta,\delta}} = \omega^{gn\delta}\ket{\theta_{\beta,\delta}}. 
\end{align}
Decomposing $\calq_g$ into two operators with support on opposite ends of the chain, the charge $(-x)$ of the left-localized operator $((\calo_{l,n})^{-x})$ is a topological invariant. Within each $(m/g)$-multiplet labelled by $\beta$, $\{\calo_{l,n},\calo_{r,n},\calq_g\}$ generate a non-commutative algebra:
\begin{align} \label{noncommutativegeneral}
\calo_{l,n}\calo_{r,n} = &\omega^{n^{\sma{2}}}\calo_{r,n}\calo_{l,n}, \;\;\;\; \calo_{l,n}\calq_g = \omega^{ng} \calq_g\calo_{l,n}, \notag \\
 &\text{and}\;\; \calo_{r,n}\calq_g = \omega^{ng}\calq_g\calo_{r,n}.
\end{align}
By our assumption of topological order ($m>g$), the phase factors $\omega^{n^2}$ and $\omega^{ng}$ are never trivially unity. For a detailed discussion of this algebra for $N=8$ and $n=2$, we direct the interested reader to App.\ \ref{app:modelcoexist}. 

\subsection{Robustness of general parafermionic phases on a ring} \label{sec:condensatestabilityclosed}

On a ring, the degeneracies due to topological edge modes are generically lost; for $g>1$, there remains a $g$-fold degeneracy on the ring due to broken symmetry. The instability of the topological degeneracy may be understood in this way: the edge modes themselves imply the existence of a $\Z_N$ singlet $\calq_g$ which splits the topological degeneracy. Suppose we couple the edges as  
\begin{align} \label{znsinglet}
\delta \calh^{\sma{(\alpha)}}=-\omega^{gn\alpha} \calq_g+h.c., \ins{with} \alpha \in \Z_{m/g}.
\end{align}
This coupling splits the open-chain $m$-multiplet into $(m/g)$ number of $g$-multiplets, as indexed by $\delta \in \Z_{\sma{m/g}}$ in Eq.\ (\ref{actionQg}). Indeed, the $m$-multiplet is a condensate of both topological edge modes and symmetry-breaking order parameters, while each $g$-multiplet is only a condensate of the symmetry-breaking order parameters. By twisting the inter-edge coupling $\delta \calh^{\sma{(\alpha)}}$ as $\alpha \rightarrow \alpha+1$, we permute the groundstate multiplet as $\delta=m/g-\alpha \rightarrow m/g-\alpha-1$.  \\

It is interesting to determine if this $g$-fold degeneracy persists as we frustrate the closed-chain Hamiltonian. Ref.\ \onlinecite{motruk2013} claims that the degeneracy is broken, while Ref.\ \onlinecite{bondesan2013} suggests that the degeneracy is robust, though without proof. The intuition for a general proof can be obtained from the simplest symmetry-broken model ($H_{\sma{(4,2)}}$) without topological order. Here, $m=n=g=2$. The $\Z_4$ symmetry is reduced to $\Z_2$ due to the parafermion order parameter $\gamma_{\sma{j}}^2$, which manifests in a two-fold-degenerate groundstate ($|\psi_{\sma{\alpha \in \Z_2,2}}\rangle$ from  Eq.\ (\ref{psin})) on both open and closed chains; we denote this groundstate projection by $P_{\sma{(4,2)}}$. We would like to know if this degeneracy is stable under deformations of $H_{\sma{(4,2)}}$, as would be implied if we prove the indistinguishability condition: $\bar{P} \bar{\calo}\bar{P} = c(\bar{\calo})\bar{P}$; here, $\bar{P}$ is a projection that is quasi-adiabatically connected to $\pft$, and $\bar{\calo}$ refers either to a perturbation, or to the Hamiltonian itself (which shall also be termed a `perturbation' in the following discussion). A legal perturbation is parafermion-local and charge-neutral; all such perturbations on an open chain are also clock-local, as proven in App.\ \ref{app:quasilocal}. We have shown in the preceding Section that the indistinguishability condition is satisfied with clock-local perturbations, thus the degeneracy is stable on an open chain. However, a closed chain allows for inter-edge perturbations which are parafermion-local but clock-nonlocal, a case in point being $\dg{\gamma}_{\sma{1}}\pdg{\gamma}_{\sma{2L}} =\omega^{\sma{3/2}}\dg{\sigma}_1\pdg{\sigma}_LQ $. What remains is to demonstrate the indistinguishability condition under clock-nonlocal perturbations, which we shall first carry out in the frustration-free limit.
A convenient intermediate step is to express $\pft$ in the charge eigenbasis: $|\phi_{\sma{\alpha,2}}\rangle = (\,|\psi_{\sma{0,2}}\rangle +(-1)^{\sma{\alpha}}|\psi_{\sma{1,2}}\rangle\,)/\sqrt{2}$. Then the off-diagonal elements of $\bar{\calm}^{\sma{(2)}}_{\sma{\ab}} = \langle \phi_{\sma{\alpha,2}}|\,\bar{\calo}\,|\phi_{\sma{\beta,2}} \rangle$ vanish because $\bar{\calo}$ is charge-neutral while different $|\phi \rangle$ carry different charge. Our frustration-free proof is complete once we demonstrate the vanishing of the difference in diagonal elements: $\bar{\calm}^{\sma{(2)}}_{\sma{00}}-\bar{\calm}^{\sma{(2)}}_{\sma{11}} = \langle \psi_{\sma{0,2}}|\,\bar{\calo}\,|\psi_{\sma{1,2}}\rangle + h.c.$. Returning to the example of $\dg{\gamma}_{\sma{1}}\pdg{\gamma}_{\sma{2L}}$, while $Q$ transforms $|\psi_{\sma{1,2}}\rangle$ into $|\psi_{\sma{0,2}}\rangle$, $\dg{\sigma}_1\pdg{\sigma}_L|\psi_{\sma{\alpha,2}}\rangle$ is orthogonal to the groundstate space. More generally, any \emph{odd} power of $\sigma_j$ (for any $j$) takes us out of the groundstate space. This follows from $\langle \psi_{\sma{\alpha,2}}| \,\sigma_{\sma{j}}^{\sma{a}}\,|\psi_{\sma{\beta,2}}\rangle =(-1)^{\sma{\alpha}}\delta_{\sma{\ab}}\delta_{\sma{a,0\,\text{mod}\,2}}$, which in turn is deducible from Eq.\ (\ref{psin}). This motivates us to try an inter-edge coupling with an even power of $\sigma_{\sma{j}}$: $\bar{\calo}=\gamma_{\sma{1}}^{2}\gamma_{\sma{2L}}^{2} \sim {\sigma}_1^2{\sigma}_L^2Q^2$. However, any \emph{even} power of $Q$ acts trivially as $\langle \psi_{\sma{\alpha,n}}|\,Q^2\,|\psi_{\sma{\beta,n}}\rangle = \delta_{\sma{\ab}}$, thus $\bar{\calo}$ cannot distinguish between $|\psi_{\sma{1,2}}\rangle$ and $|\psi_{\sma{0,2}}\rangle$. Finding a parafermion-local operator that would do so turns out to be impossible, as we now show. Any inter-edge coupling may be decomposed into the form: $B_lB_rQ^{[[B_r]]}$, where $B_l$ (resp.\ $B_r$) is a clock-local operator with definite charge, and with support near site $1$ (resp.\ $L$). The Jordan-Wigner transformation\ (\ref{pfclocktransformation}) implies that there are as many powers of the string $Q$ as the charge of $B_r$. To transform $|\psi_{\sma{1,2}}\rangle$ into $|\psi_{\sma{0,2}}\rangle$, one needs an odd power of the string operator $Q$; however, an odd power of $Q$ always accompanies an odd power of $\sigma_j$ (in $B_r$), which takes us out of the groundstate space. This intuitive observation in the frustration-free limit can be supplemented with quasi-adiabatic techniques to prove the indistinguishability condition in the presence of frustration, as we show in App.\ \ref{app:indistpflocal}.\\

We now extend our discussion to more general $(N,n)$-phases with a parafermionic order parameter, possibly in conjunction with topologically order. Let us decompose the groundstate space $P[H_{\sma{op}}]$ of an open-chain Hamiltonian into eigenstates of the $\Z_N$ generator: each of $|\phi^s_{\sma{\alpha \in \Z_m,n}}\rangle$ has a conserved charge $n\alpha$. For $g>1$, the parafermionic order parameter $\gamma_{\sma{j}}^{\sma{N/g}}$ (or its dressed version $\caly_j$) permutes the groundstates as $|\phi^s_{\sma{\alpha,n}}\rangle \rightarrow |\phi^s_{\sma{\alpha-m/g,n}}\rangle$, as evidenced from $\gamma_{\sma{j}}^{\sma{N/g}}Q = {\omega}^{\sma{N/g}}Q\gamma_{\sma{j}}^{\sma{N/g}}$. Thus $P[H_{\sma{op}}]$ divides into $m/g$ number of parafermion-condensed subspaces, which we label by $\delta \in \Z_{m/g}$ and denote as 
\begin{align}
\bar{P}_{\delta}=\sum_{\alpha=0}^{g-1}\ket{\phi^s_{\alpha m/g+\delta,n}}\bra{\phi^s_{\alpha m/g+\delta,n}}.
\end{align}
We propose that each of $\bar{P}_{\delta}$ is indistinguishable under any symmetric, \emph{parafermion}-local probe $\bar{\calo}$, i.e., in the thermodynamic limit,
\begin{align} \label{spostrong}
\bar{P}_{\delta}\bar{\calo}\bar{P}_{\delta} = \bar{c}(\bar{\calo}) \bar{P}_{\delta}
\end{align}
for a complex number $\bar{c}$. The proof of Eq.\ (\ref{spostrong}) generalizes our intuition from the $H_{\sma{(4,2)}}$ model, and may be found in App.\ \ref{app:indistpflocal}. We note that Eq.\ (\ref{spostrong}) is a stronger statement than the local-indistinguishability condition\ (\ref{spo}), which applies \emph{only} to symmetric, \emph{clock}-local probes of $P[H_{\sma{op}}]$. By `stronger' indistinguishability, we mean that a certain groundstate space ($\bar{P}_{\delta}$) cannot be distinguished by both clock-local and -nonlocal operators (which are symmetric and parafermion-local). Suppose we now close the chain symmetrically; any legal term in the closed-chain parafermion Hamiltonian satisfies the same conditions on $\bar{\calo}$. Since each of $\bar{P}_{\delta}$ satisfies the strong indistinguishability condition, but their direct sum $P[H_{\sma{op}}]=\sum_{\delta=0}^{m/g-1}\bar{P}_{\delta}$ does not, the open-chain groundstate splits into multiplets. A multiplet indexed by $\delta$ is robustly degenerate, so long as the gap persists within each of the $g$ charge sectors of $\bar{P}_{\delta}$.


\section{Discussion} \label{sec:discussion}

Topological order is known to characterize the low-energy subspace for a variety of condensed-matter systems.\cite{kitaevPreskill2006,LevinWen2006,li2008,aa2011,turner2011,motruk2013} In this paper, we have precisely characterized the groundstate of 1D parafermionic chains, and identified the properties which are robustly associated with topological order and/or symmetry-breaking. Our work generalizes a previous work that imposed constraints on the high-energy states.\cite{fendley2012} These constraints cannot be realized in nature, since parafermions only emerge as low-energy quasiparticles.\cite{clarke2013,lindner2012,mengcheng2012} \\

A unifying property of topologically-ordered groundstates is their indistinguishability by local probes. From this fundamental property, one may derive:\cite{hastingsLesHouches} (i) a nontrivial groundstate degeneracy depending on the topology of the manifold, and (ii) well-known signatures in the entanglement entropy.\cite{kitaevPreskill2006,LevinWen2006} Besides parafermions, a close variant\cite{brayvi2010,brayvi2011} of Eq.\ (\ref{spo}) applies to other models with topological order, including the toric code\cite{kitaev2003} and Levin-Wen string-net models;\cite{levinwen2005} a stronger version of Eq.\ (\ref{spo}) is known to stabilize the spectral gap of frustration-free Hamiltonians under generic local perturbations, and produces an area-law for the entanglement entropy.\cite{michalakis2013}\\

A few works\cite{fidkowski2011b,ortiz2014} have demonstrated that Majorana edge modes may exist with weaker localization properties in particle-number-conserving superconductors, where the $U(1)$ symmetry (associated with the electron charge) is not spontaneously broken to $\Z_2$. For one particular soluble model, it has been shown that the groundstate of the topological phase switches parity when the boundary condition is changed;\cite{ortiz2014} it is interesting to determine if this property is more generally true away from the soluble limit, perhaps with similar techniques that are presented in this paper. \\

In the final stages of this work, the phase diagram for our $\Z_3$ model has been alternatively derived from an entanglement perspective in Ref.\ \onlinecite{yezhuang2015}. \\

\noindent \emph{Acknowledgements:} The authors thank Paul Fendley, David Huse, Elliot Lieb and Kim Hyungwon for their expert opinions on various subjects in this paper. AA also acknowledges discussions with Jason Alicea, Roger Mong, Roman Lutchyn, Yeje Park, Yang-Le Wu, Jian Li, Curt von Keyserlingk, Titus Neupert, Dan Arovas and Zhoushen Huang. NR and CF acknowledges P. Lecheminant for fruitful discussions about the XYZ model. AA and BAB were supported by NSF CAREER DMR-095242, ONR - N00014\text{-}11\text{-}1-0635, MURI\text{-}130\text{-}6082,  NSF-MRSEC DMR-0819860, Packard Foundation and Keck grant. CF and MJG were supported by the Office of Naval Research under grant N0014-11-1-0123. NR was supported by the Princeton Global Scholarship. MJG was supported by the NSF CAREER EECS-1351871. This work was also supported by DARPA under SPAWAR Grant No.: N66001\text{-}11\text{-}1-4110.


\begin{widetext}

\appendix

\section{Local operators and quasi-adiabatic continuation} \label{app:quasilocal}

To define a local operator ($\calo$), we first decompose it as 
\begin{align} \label{decomposition}
\calo = \sum_{r \geq 1} \sum_{A \in S(r)} V_{r,A},
\end{align}
where $V_{r,A}$ acts nontrivially on the chain $A$ of length $r$, and $\cals(r)$ is the set of all chains with length $r$. For a chain of length $L$, $\calo$ acts in the Hilbert space: $\calz = (\mathbb{C}^{N})^{\otimes L}$. The operator norm $||\call||_{\sma{\text{op}}}$ of an operator $\call$ is defined by the supremum of $\{ || \call v||: v \in \calz$ with $||v||=1\}$. We call $\calo$ quasilocal if it can be characterized by strength $K$ and decay $f(r)$, such that the operator norm $|| V_{r,A}||_{\sma{\text{op}}} \leq K f(r)$ for all $r \geq 1$ and for any chain ($A$) of length $r$; $f: \Z_+ \mapsto [0,1]$ and decays superpolynomially in $r$, i.e., faster than any power. Included are finite-ranged and  exponentially-decaying interactions. For example, the strength of the finite-ranged Hamiltonian\ (\ref{pfHamn1}) is $J$. In this example and the rest of the paper, we consider only quasilocal Hamiltonians, which are characterized by a finite Lieb-Robinson velocity,\cite{lieb1972} i.e., information effectively propagates at finite speed, in analogy with the speed of light. This implies that any finite-time evolution by a quasilocal Hamiltonian preserves the locality of an operator.\cite{hastings2010} We now elaborate on an important example of such an evolution: a quasi-adiabatic continuation. For illustration, let us symmetrically deform $\hno$ of Eq.\ (\ref{pfHamn1}), while preserving the spectral gap above the lowest $N$ states (which we do not assume to be degenerate). We assume there exists a family of symmetry-preserving Hamiltonians $H_s$, which are differentiable in $s$ for $s \in [0,1]$, and  $H_0 \equiv \hno$.  The spectral gap allows us to uniquely define $P_s$ as the projection into the lowest $N$ states of $H_s$. We define an exact quasi-adiabatic continuation $\calv_s$ that maps ${P}_s = \calv_s {P}_0 \dg{\calv}_s$;\cite{hastings2005} $\calv_s$ is the unitary transformation\ (\ref{qqq3}) with the Hermitian Hamiltonian\ (\ref{qadiab3}). The infinite-time evolution in Eq.\ (\ref{qadiab3}) is filtered by a function $F(t)$ that decays faster than any power, for large $|t|$; an explicit construction of $F(t)$ is provided in the appendices of Ref.\ \onlinecite{brayvi2011} and\ \onlinecite{hastings2010}. We assume for simplicity that the interactions in $H_s$ decay exponentially, or faster. It follows from Lemma 2 of Ref.\ \onlinecite{brayvi2011} that $\cald_s$ is a quasilocal Hamiltonian. By interpreting $s \in [0,1]$ as a time variable, Lemma 1 of Ref.\ \onlinecite{brayvi2011} informs us that the finite-time evolution $\calv_s$ generated by $\cald_s$ preserves locality, i.e., $\dg{\calv}_s \calo {V}_s$ is a quasilocal if $\calo$ is quasilocal. Since $[H_s,Q]=0$, it follows that the quasi-adiabatic continuation is symmetry-preserving, i.e., $[\cald_s,Q]=[\calv_s,Q]=0$.\\

It is useful to distinguish between two notions of locality on a closed chain. We say that an operator is clock-local (parafermion-local) if it has strength $J$ and superpolynomial decay $f(r)$ in the clock (parafermion) representation. We describe parafermions (clocks) only with a parafermion-local (clock-local) Hamiltonian. We also insist that all Hamiltonian terms are charge neutral, by which we mean they commute with the $\Z_N$ generator. On an open chain, we now demonstrate that all parafermion-local terms are also clock-local, due to the charge-neutral constraint. Decomposing a generic parafermion-local term as in Eq.\ (\ref{decomposition}), we consider each $V_{r,A}$ separately: 
\begin{align}
V_{r,A} \propto \gamma_i^{n_i}\gamma_{i+1}^{n_{i+1}} \ldots \gamma_{i+r-1}^{n_{i+r-1}},
\end{align}
where $n_j \in \Z_N$, and by assumption of an open chain, $i \geq 1$ and $i+r-1 \leq L$. Let us express this term in the clock representation through Eq.\ (\ref{pfclocktransformation}). Loosely speaking, this transformation introduces strings of $\tau_j$ for $j<i$, assuming $i>1$. However, if $V_{r,A}$ is charge neutral, the strings all cancel out. More precisely, we operate on each site labelled by $j<i$ as $\tau_j^{\bar{n}}$ with $\bar{n}={\sum_{k=i}^{i+r-1}n_k}$. However,
\begin{align}
\dg{Q}V_{r,A}Q = \omega^{\bar{n}} V_{r,A},
\end{align}
and charge neutrality imposes that $[Q,V_{r,A}]=0$, or equivalently, $\bar{n} =0 $ mod $N$. \\

The discussion thus far applies to an open chain. On a closed chain, charge-neutral inter-edge couplings can be either (i) parafermion-local but clock-nonlocal (e.g. Eq.\ (\ref{examplesupport})), or (ii) clock-local but parafermion-nonlocal (e.g. $\dg{\sigma}_1\pdg{\sigma}_L$). Refining our notion of locality thus distinguishes between parafermions on a ring and clocks on a ring.

\section{Indistinguishability by general quasilocal probes} \label{app:generalprooflocalindist}

In Sec.\ \ref{sec:localindis}, we claimed that (a) $\pno$ satisfies the local-indistinguishability condition\ (\ref{spo}), for any $\Z_N$-symmetry-preserving, clock-local operator $\calo$. More precisely, $\calo$ has strength $K$ and superpolynomial decay $f(r)$ in the clock representation, as we describe in App.\ \ref{app:quasilocal}. Later in Sec.\ \ref{sec:reviewpfgeneraln}, we also claimed that (b) $\pnn$ is locally indistinguishable. The proofs of both statements can be combined. In what follows, $n$ is a divisor of $N$, and $m=N/n$. It suffices to show that the matrix $\calm^{(n)}_{\ab}=\bra{\phi_{\alpha,n}}\calo\ket{\phi_{\beta,n}}$ is proportional to the identity, where $\ket{\phi_{\sma{\alpha,n}}} \in P[\hnn]$ is defined in Eq.\ (\ref{chargeeigenbasis}). The off-diagonal elements vanish because $[\calo,Q]=0$ by assumption. The difference in diagonal elements is
\begin{align} \label{diffdiag}
\calm^{(n)}_{\alpha \alpha}- \calm^{(n)}_{\beta \beta}=\sum_{\mu,\nu \in \Z_m; \mu \neq \nu}\frac{\omega^{\alpha n (\mu-\nu)}-\omega^{\beta n(\mu-\nu)}}{m}\bra{\psi_{\mu,n}} \calo \ket{\psi_{\nu,n}}.
\end{align}
We claim that 
\begin{align}
\calm^{(n)}_{\alpha \alpha}- \calm^{(n)}_{\beta \beta} \leq 2(m-1)Kf(L)
\end{align}
for a chain of length $L$.\\

\noindent \emph{Proof}: For $\calo$ to have nonvanishing matrix elements between $\bra{\psi_{\mu,n}} \calo \ket{\psi_{\nu,n}}$ for $\mu\neq \nu$, we require that $\calo$ flips the clock variable on all $L$ sites. Thus if we decompose $\calo$ as in App.\ \ref{app:quasilocal}, we find $\bra{\psi_{\mu,n}} \calo \ket{\psi_{\nu,n}} = \bra{\psi_{\mu,n}} V_{L,A} \ket{\psi_{\nu,n}}$, where $V_{L,A}$ acts nontrivially on the entire chain. By the Cauchy-Schwartz inequality,
\begin{align}
\bigg|\bra{\psi_{\mu,n}}  V_{L,A} \ket{\psi_{\nu,n}}\bigg| \leq  \bigg|\bigg| \ket{\psi_{\mu,n}}\bigg|\bigg| \cdot \bigg|\bigg|V_{L,A} \ket{\psi_{\nu,n}}\bigg|\bigg| = \bigg|\bigg|V_{L,A} \ket{\psi_{\nu,n}}\bigg|\bigg|.
\end{align}
Now applying the definitions of the operator norm and quasilocality (cf. App.\ \ref{app:quasilocal}),
\begin{align}
\bigg|\bigg|V_{L,A} \ket{\psi_{\nu,n}}\bigg|\bigg| \leq \bigg|\bigg|V_{L,A}\bigg|\bigg|_{\sma{\text{op}}} \leq K\,f(L).
\end{align}
It follows from this and Eq.\ (\ref{diffdiag}) that
\begin{align}
\bigg|\bra{\phi_{\alpha,n}}\calo\ket{\phi_{\alpha,n}} - \bra{\phi_{\beta,n}}\calo\ket{\phi_{\beta,n}}\bigg| \leq \frac{2}{m}\sum_{\mu,\nu \in \Z_m; \mu \neq \nu}\bigg|\;\bra{\psi_{\mu,n}} \calo \ket{\psi_{\nu,n}}\;\bigg| \leq 2(m-1)\,K\,f(L).
\end{align}

\section{Generalized edge mode for the $\Z_3$ parafermion model} \label{app:edgemode}

Let us evaluate the zero edge mode $\calo_l$ for the $\Z_3$ model\ (\ref{Z3example}), which we rewrite as $H_s=H_0+sV$, with $s=f/J$,
\begin{align}
H_0 = -\hat{\lambda}\sum_{j=1}^{L-1}\omega \dg{\gamma}_{\sma{2j+1}}\pdg{\gamma}_{\sma{2j}} + h.c., \;\hat{\lambda}=e^{i\hat{\phi}}, \ins{and} V = -\sum_{j=1}^L\omega^*\dg{\gamma}_{\sma{2j-1}} \pdg{\gamma}_{\sma{2j}} + h.c..
\end{align}  
We need certain relations introduced in Sec.\ \ref{sec:zeromodes}, including the definitions of the quasi-adiabatic continuation operator $\calv_s$ in Eq.\ (\ref{qqq3}), which is generated by the Hamiltonian $\cald_s$ in Eq.\ (\ref{qadiab3}). The form of $\calo_l$ in Eq.\ (\ref{edgemodefirstorder}) motivates us to evaluate the operator $i[\cald_0,\gamma_{\sma{1}}]$. Employing $[H_0,\go]=0$ and $[V,\go]= ({\omega}^*-\omega)(\gt -\omega \dg{\go}\dg{\gt})$, we are led to
\begin{align} \label{caldgam1}
i[\cald_0,\go] = (\omega^*-\omega)(X-\omega \dg{\go}Y) 
\end{align}
\begin{align}
\ins{with} X = \int dt F(3 t) \, e^{iH_0t}\,\gt\,e^{-iH_0t} \ins{and} Y =  \int dt F(3 t) \, e^{iH_0t}\,\dg{\gt}\,e^{-iH_0t}.
\end{align}
Here, we have applied that the spectral gap $\Gamma^0$ (above the lowest three eigenstates of $H_0$) equals $3$. The problem is reduced by noting $Y=-\dg{X}$, as follows from $F$ being imaginary. Further progress is made by realizing that $\gt,\gth$ and $\omega \dg{\gt}\dg{\gth}$ form a closed linear algebra under commutation by $H_0$. That is, the coefficients defined by
\begin{align}
\big[ \,H_0,\,A_0\gt + B_0\gth+ C_0\,\omega \dg{\gt}\dg{\gth}\,\big]= A_1\gt + B_1\gth+ C_1\, \omega \dg{\gt}\dg{\gth}
\end{align}
are linearly related as
\begin{align} \label{definecalm}
\begin{pmatrix} A_1 \\ B_1\\ C_1 \end{pmatrix} = 3\calm \begin{pmatrix} A_0 \\ B_0\\ C_0 \end{pmatrix} \ins{with Hermitian matrix} \calm = \frac{1}{i\sqrt{3}}  \begin{pmatrix} 0 & -\hat{\lambda} & \hat{\lambda}^* \\ \hat{\lambda}^* & 0 & -\hat{\lambda} \\ -\hat{\lambda} & \hat{\lambda}^* & 0 \end{pmatrix}.
\end{align}
Diagonalizing $\calm$, we find eigenoperators:
\begin{align} \label{definevj}
v_j = \oneover{\sqrt{3}}( \gt + \omega^j\gth + \omega^{2j+1} \dg{\gt}\dg{\gth}), \ins{with corresponding eigenvalues} \zeta_j(\hat{\phi}) = -\frac{2}{\sqrt{3}} \si \big(\hat{\phi} + \tfrac{2\pi}{3}j\big),
\end{align}
such that $[H_0,v_j]=3\zeta_jv_j$. Expanding $\gt$ in this eigenbasis as $\gt = (v_0+v_1+v_2)/\sqrt{3}$.
\begin{align} \label{defineX}
X = \frac{1}{\sqrt{3}}  \int dt F(3t) \sum_{j=0}^2 v_j e^{i3 \zeta_j t}  =  \frac{1}{3\sqrt{3}}\sum_{j=1}^3 \tilde{F}(\zeta_j) v_j.
\end{align}  
To proceed, we thus need the Fourier components $\tilde{F}$ for the frequencies $\pm \zeta_j$. How are they determined? The two essential properties of a quasi-adiabatic continuation $\calv_s$ is that (a) it evolves the groundstate space $P_s$ as a function of the deformation parameter $s$, and (b) that it is locality-preserving. The first property (a) determines the Fourier components $\tilde{F}(\Omega)=-1/\Omega$ for $|\Omega|\geq 1$ (as shown in Sec.\ \ref{sec:zeromodes}), while the second (b) determines the components for $|\Omega|<1$, up to a degree of arbitrariness. As will shortly be clarified, this arbitrariness lies in the operators which have zero matrix elements within the groundstate space. Now we address separately the chiral and nonchiral cases: the nonchiral limit $\hat{\phi}=0$ is discussed in App.\ \ref{phizero}, and  $0<\hat{\phi}<\pi/3$ is elaborated in App.\ \ref{arbitrary}.
 
\subsection{Nonchiral $\Z_3$ parafermions: $\hat{\phi}=0$} \label{phizero}

As defined in Eq.\ (\ref{definevj}), the eigenvalues of $\calm$ are $\zeta_0 =0, \zeta_1 =-1$ and $\zeta_2=1$. For the first eigenvalue, the oddness of $\tilde{F}$ implies that $\tilde{F}(\zeta_0)=0$; for $j \in \{1,2\}$, $\tilde{F}(\zeta_j) = -1/\zeta_j$ since $|\zeta_j| \geq 1$. Eq.\ (\ref{defineX}) leads to
\begin{align}
X =  \frac{1}{3\sqrt{3}}( v_1-v_2)  = \frac{1}{3(\omega^*-\omega)} \big( \gamma_3 - \omega \dg{\gamma}_2\dg{\gamma}_3),
\end{align}
Inserting the expressions of $X$ and $Y=-\dg{X}$ into Eq.\ (\ref{caldgam1}), we thus derive from Eq.\ (\ref{edgemodefirstorder}) the form of $\calo_l$, to first order in the deformation parameter $s$. This result is found in Eq.\ (\ref{edgemodeZ3}).

\subsection{Chiral $\Z_3$ parafermions: $0<\hat{\phi}<\pi/3$} \label{arbitrary}

In this range of $\hat{\phi}$, both $|\zeta_0|$ and $|\zeta_1|<1$, while $|\zeta_2|>1$. To determine the edge mode, we thus need the Fourier component $\tilde{F}(\Omega)$ for $0<|\Omega|<1$, which enters the expression\ (\ref{defineX}). In the conventional quasi-adiabatic continuation, $\tilde{F}$ in this range is optimized so that $\calv_s$ {maximally} preserves locality, which amounts to finding $F(t)$ with the fastest possible decay. This has been proven to be of subexponential type,\cite{hastings2010,Ingham} that is, $F(t)$ decays faster than exp$-t^{\alpha}$ for any $\alpha <1$; an example of $F$ is described in the next paragraph. We refer to subexponentially-decaying edge modes ($\calo_l$) as maximally localized. On the other hand, one may choose an edge mode $\bar{\calo}_l$ with worse localization properties, but with the desirable property that it commutes with the Hamiltonian. There is thus a trade-off between localizability and commutativity. Indeed, choosing $\tilde{F}(\Omega)=-1/\Omega$ for $|\Omega|=|\zeta_i|<1$ worsens the decay of $F(t)$ (and thus of the resultant edge mode), but leads to an edge mode which commutes with the Hamiltonian, as we elaborate in App.\ \ref{app:caseb}. One may verify that the difference $\Delta = \bar{\calo}_l-\calo_l$ has zero matrix elements in the groundstate space. To illustrate this for the $\Z_3$ chiral example, we have shown that the first-order correction to the edge mode has a term proportional to $\sum_{j=1}^3 \tilde{F}(\zeta_j) v_j$; this follows from Eq.\ (\ref{caldgam1}) and (\ref{defineX}). Since $|\zeta_0|,|\zeta_1|<1$, the coefficients of $v_0,v_1$ (as defined in Eq.\ (\ref{definevj})) indeterminately depend on our choice of the filter function in the range $0<|\Omega|<1$; once again, we are choosing between spatial localization or commutivity with the full Hamiltonian. This lack of determinacy has no physical consequence in the groundstate space, as $v_j \ket{\psi_{\alpha,1}}=0$ for $j \in \{0,1\}$ and $\alpha \in \Z_3$; here, recall $\psi$ are the groundstates of $H_0$, as defined in Sec.\ \ref{sec:localindis}.\\

To aid future numerical studies, we provide further details of the optimized filter function. Here, we do not claim originality, but hope to present an accessible summary of certain portions in Ref.\ \onlinecite{hastings2010} and \onlinecite{Ingham}. Expressing
\bal
\tilde{F}(\Omega) = -\frac{1-\tilde{g}(\Omega)}{\Omega},
\end{align}
we define a real, even function $\tilde{g}$ to satisfy: (i) $\tilde{g}(0)=1$, and (ii) $\tilde{g}$ has compact support within the interval $[-1,1]$. The uncertainty principle then states that $g$ cannot be `very small' -- optimally,\cite{Ingham} $g(t) \leq C$ exp$[-|t|\varepsilon(|t|)]$  for some constant $C,$ and $\varepsilon$ a positive, monotone-decreasing function satisfying that
\bal
\int^{\infty}_1 \frac{\varepsilon(y)}{y} dy \ins{converges.}
\end{align}
It is further shown by Ingham in Ref.\ \onlinecite{Ingham} how to construct $g$ such that
\bal
\varepsilon(t) = \frac{1}{(\text{log}(2+t))^2},
\end{align}
though it has been noted that faster asymptotic decays with different $\varepsilon$ are possible in principle.\cite{hastings2010} For Ingham's choice,
\bal
g(t) = A \prod_{n=1}^{\infty} \frac{\si (\rho_n t)}{\rho_nt},
\end{align}
with 
\bal \label{rhon}
\rho_n = \begin{cases} \frac{B \varepsilon(n_0)}{n_0}, & n \leq n_0  \\   \frac{B \varepsilon(n)}{n}, & n > n_0. \end{cases}
\end{align}
$A$, $B$ and $n_0$ are chosen such that $\tilde{g}$ satisfies the above properties (i) and (ii); the final result is shown in blue in Fig.\ \ref{fig:filter}.\\

\begin{figure}[H]
\centering
\includegraphics[width=6 cm]{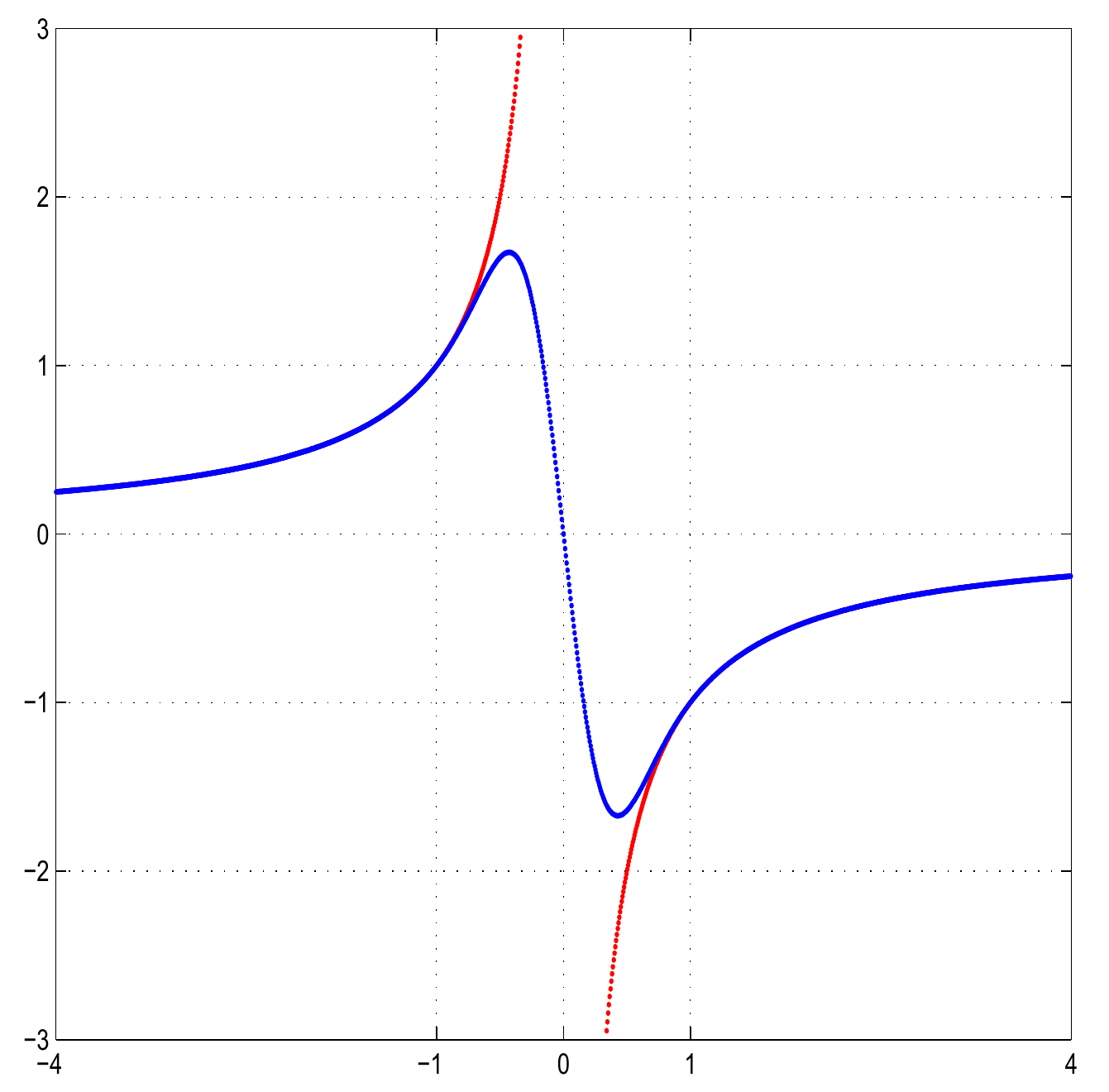}
\caption{ Blue: optimized filter function $(\tilde{F}(\Omega))$ obtained by roughly estimating the parameters in Eq.\ (\ref{rhon}) as $A = 1/9.2843, \, B =  80, \, n_0=100$. Red: $-1/\Omega$ is shown for comparison. }\label{fig:filter}
\end{figure}

\section{Zero edge modes that commute with the Hamiltonian} \label{app:caseb}

Let us define the first-order truncation of $\calo_l$ as $\calo_l^{\sma{(1)}} = \gamma_1 + is[\cald_0,\gamma_1],$ where $s$ is the deformation parameter. This operator depends on our choice of the filter function $F$, which enters the expression for $\cald_s$ in Eq.\ (\ref{qadiab3}). For a specific choice of $F$ (with corresponding edge mode $\bar{\calo}_l^{\sma{(1)}}$), we find that $\bar{\calo}_l^{\sma{(1)}}$ and $H_s$ commute to first order in $s$. The following discussion clarifies this choice of $F$. Let $v_j$ be operators which satisfy two properties: (i) they belong in the charge sector $\alpha$, i.e., $\dg{Q}v_jQ=\omega^{\alpha}v_j$, and (ii) they are constants under time evolution by $H_0$, i.e., $[H_0,v_j]= 2\zeta_jv_j$ with real eigenvalues $2\zeta_j$. These operators $v_j$ span the space ($\calx_{\alpha}$) of operators  in the same charge sector, i.e., for each element $x \in \calx_{\alpha}$, $\dg{Q}xQ=\omega^{\alpha}x$. Labelling each element by a subscript (e.g., $x_i$), commutation by $H_0$ may be interpreted as an eigenvalue problem: $[H_0,x_i]= \calb_{ij}x_j$. The matrix $\calb$ is known to be Hermitian in the frustration-free limit,\cite{fendley2012} where $H_0=H_{N,1}$, as described in Eq.\ (\ref{pfHamn1}). Applying the completeness property, we expand $[V,\gamma_1] = \sum_n r_n v_n$, with c-numbers $r_n$. This leads to
\begin{align}
i[\cald_0,\gamma_1] = \sum_n r_n \int dt F(\Gamma_0t) e^{iH_0t}v_n e^{-iH_0t} = \sum_n \frac{r_n v_n}{\Gamma_0} \tilde{F}(2\zeta_n/\Gamma_0).
\end{align}
If the Fourier transform $\tilde{F}(\Omega)$ is chosen as $-1/\Omega$ for all relevant frequencies $\Omega \in \{2\zeta_n/\Gamma_0\}$ in the above sum, we denote the resultant edge mode by $\bar{\calo}_l$. This choice of $\tilde{F}$, which we call Fendley's choice, is only possible when none of the relevant frequencies are vanishing, i.e., $\zeta_n \neq 0$ for any nonzero $r_n$. The proof of commutivity to first order is elementary:  
\begin{align}
[H_s,\bar{\calo}^{(1)}_l] =&\; s[V,\gamma_1]+s[H_0,i[\cald_0,\gamma_1]] + O(s^2) = s[V,\gamma_1] +s \sum_n \frac{r_n [H_0,v_n]}{\Gamma_0} \tilde{F}(2\zeta_n/\Gamma_0) + O(s^2)  \notag \\
=&\; s[V,\gamma_1] +s \sum_n \frac{2 r_n \zeta_n v_n}{\Gamma_0} \bigg(-\frac{\Gamma_0}{2\zeta_n}\bigg) + O(s^2) = O(s^2).
\end{align}
In cases where all the relevant frequencies are large, by which we mean they satisfy: $|2\zeta_n/\Gamma_0| \geq 1$, Fendley's choice is not so much a choice but a necessity. Indeed, Fendley's choice then coincides with Eq.\ (\ref{FTfilter}) in Sec.\ \ref{sec:zeromodes}; we have shown therein that Eq.\ (\ref{FTfilter}) leads to the correct evolution of the groundstate space: $\partial_sP_s = i[\cald_s,P_s]$. The set of cases where all the relevant frequencies are large include all quadratic Majorana models, as we show in App.\ \ref{app:edgemodeZ2}. On the other hand, if some of the relevant frequencies satisfy: $|2\zeta_n/\Gamma_0|<1$, Fendley's choice produces an edge mode that is deformed from the maximally-localized edge mode. In App.\ \ref{app:z3nonchiralfermion}, we carry out this deformation for the $\Z_3$ chiral parafermion.

\subsection{Zero edge mode for quadratic Majorana models} \label{app:edgemodeZ2}

Suppose we deform the topological dimer model described in Eq.\ (\ref{frustratedZ2model}). If we limit our deformations to terms which are quadratic in the Majorana operators, we can derive a general expression for the zero edge mode in the first-order approximation. To begin, let us express the dimer model ($H_0$) and the deformation ($V$) as:
\begin{align}
H_0 =\frac{i}{2} \gamma_i A_{ij} \gamma_j,\;\; V =  \frac{i}{2} \gamma_i B_{ij} \gamma_j,
\end{align}
with $A$ and $B$ real, skew-symmetric, and even-dimensional matrices. It follows that the eigenvalues of $A$ (and $B$) are completely imaginary, and come in complex-conjugate pairs. The eigenproblem $[H_0,\gamma_j] = -2i \sum_m A_{jm}\gamma_m$ thus has real eigenvalues $\zeta_n$: $[H_0,v_n]=2\zeta_nv_n$. The eigenvectors $v_n$ form a complete basis for all linear Majorana operators, thus we may define the expansion $\gamma_j = \sum_l z_{jl} v_l$; alternatively, $z_{jl}$ may be obtained from diagonalizing $A$. The eigenvalues $\{\zeta_j\}$ of $-iA$ are interpreted as half the single-particle energies of $H_0$, i.e., $H_0$ has single-particle excitations with energy $2\zeta_j$. Since $H_0$ is the topological dimer model described in Eq.\ (\ref{frustratedZ2model}), we know that $A$ has two zero modes (denoted $\zeta_1=\zeta_2=0$), corresponding to the edge operators $v_1=\gamma_1$ and $v_2=\gamma_{2L}$. All other single-particle excitations are gapped, i.e.,  $|2\zeta_j| \geq \Gamma_0$ for $j >2$, where $\Gamma_0$ is the many-body spectral gap of $H_0$. The edge mode is derived as  
\begin{align}
i[\cald_0,\gamma_1] =&\;  \int dt F(\Gamma_0t)e^{iH_0t}[V,\gamma_1]e^{-iH_0t} = -2i \sum_{m \neq 1,2L} B_{1m} \int dt F(\Gamma_0t)e^{iH_0t}\gamma_m e^{-iH_0t} \notag \\
=&\; -2i  \sum_{m \neq 1,2L}\sum_{n \neq 1,2} B_{1m}z_{mn} v_n \int dt F(\Gamma_0t)e^{i2\zeta_nt} = -2i \Gamma_0^{\mo} \sum_{m \neq 1,2L}\sum_{n \neq 1,2} B_{1m}z_{mn} v_n \tilde{F}(2\zeta_n /\Gamma_0).
\end{align}
Since $V$ is skew-symmetric, the sum over $m$ excludes $1$; this sum also excludes $2L$, since $\gamma_{2L}$ is a zero mode and $F$ an odd function (recall Sec.\ \ref{sec:zeromodes}). Since $H_0$ does not contain a term proportional to either $\gamma_1$ or $\gamma_{2L}$, it follows that the expansion $\gamma_{m \neq 1,2L} =  \sum_{n \neq 1,2} z_{mn} v_n$ excludes the zero modes $v_1=\gamma_1$ and $v_2=\gamma_{2L}$. Alternatively stated, all the relevant frequencies are gapped, i.e., $|2\zeta_n /\Gamma_0| \geq 1$. Now applying that $\tilde{F}(\Omega) = -1/\Omega$ for these relevant frequencies,  we find
\begin{align}
i[\cald_0,\gamma_1] =&\;  i \sum_{m \neq 1,2L}\sum_{n \neq 1,2} B_{1m}z_{mn} \frac{v_n}{\zeta_n}.
\end{align}
Applying this to the quadratic Majorana model\ (\ref{frustratedZ2model}), we obtain $i[\cald_0,\gamma_1]=\gamma_3$, as we have previously derived in less-general fashion; cf. Sec.\ \ref{sec:zeromodes}.

\subsection{Deformed zero edge mode for the $\Z_3$ chiral parafermion} \label{app:z3nonchiralfermion}

For illustration, we consider the $\Z_3$ chiral parafermion with $0<\hat{\phi}<\pi/3$. Recall that the filter function has not been specified in Eq.\ (\ref{defineX}). If we choose $\tilde{F}(\zeta_j) = -1/\zeta_j$, we alternately derive Fendley's edge mode. Substituting this choice into Eq.\ (\ref{defineX}) and applying Eq.\ (\ref{definevj}), we obtain
\begin{align}
\bar{X}=-\frac{1}{3\sqrt{3}} \sum_{j=0}^2 \frac{ v_j}{{\zeta}_j} = -\frac{1}{9}\bigg[\big(\frac{1}{\zz} + \frac{1}{\zo} +\frac{1}{\zt}\big)\gt + \big(\frac{1}{\zz} + \frac{\omega}{\zo} +\frac{\omega^*}{\zt} \big)\gth +\big( \frac{1}{\zz} + \frac{\omega^*}{\zo} +\frac{{\omega}}{\zt}\big)\omega \dg{\gt}\dg{\gth}\bigg].
\end{align}
Here, we have introduced $\bar{X}$ (to distinguish Fendley's choice) in place of $X$ (which we reserve for the conventional quasi-adiabatic continuation). A useful relation is $\zeta_1\zeta_2\zeta_3 = 2\si(3\hat{\phi})/(3\sqrt{3})$, which is derived from taking the determinant of $\calm$ as defined in Eq.\ (\ref{definecalm}). This relation, in combination with elementary trigonometric identities, leads to
\begin{align}
(\omega^*-\omega)\bar{X} = -\frac{i}{2 \si (3\phi)} \big( \gamma_2 + \hat{\lambda}^2\gamma_3 + \hat{\lambda}^{-2}\omega \dg{\gamma}_2\dg{\gamma}_3 \big).
\end{align}
The expressions for $\bar{X}$ and $\bar{Y}=-\dg{{\bar{X}}}$, in combination with Eq.\ (\ref{caldgam1}) and\ (\ref{edgemodefirstorder}), lead to the final result
\begin{align} \label{edgemodecommutes}
\bar{\calo}_l = \gamma_1 -\frac{is}{2 \si (3\phi)} \bigg[ \gamma_2 + \hat{\lambda}^2\gamma_3 + \hat{\lambda}^{-2}\omega \dg{\gamma}_2\dg{\gamma}_3 +\omega \dg{\gamma_1}\big(\dg{\gamma}_2 +\hat{\lambda}^{-2}  \dg{\gamma}_3 +\hat{\lambda}^2 \omega {\gamma}_2 {\gamma}_3\big) \bigg] + \ldots
\end{align}
This deformed edge mode $\bar{\calo}_l$ is identical to the expression $\Psi_{\sma{\text{left}}}$ alternatively derived in Ref.\ \onlinecite{fendley2012}, up to minor typographic errors in that reference. At least to first order in $s$, Fendley's edge mode can thus be understood as a deformed quasi-adiabatic continuation of $\gamma_1$; the resultant operator is not optimally localized. This deformation cannot always be justified: as $\hat{\phi}$ (and $\zeta_0$) tends to zero, the localization of $\bar{\calo}_l$ worsens dramatically, as evidenced by the diverging first-order coefficient in Eq.\ (\ref{edgemodecommutes}), and the increasingly 
singular $\tilde{F}$. In contrast, the conventional quasi-adiabatic continuation chooses a filter function that is everywhere infinitely differentiable;\cite{hastings2010} the resultant edge mode is well-defined for any $\hat{\phi}$, as shown in App.\ \ref{app:edgemode}.

\section{Closing the chain by imposing translational invariance} \label{app:twistmethod}

Our goal is to map a parafermion Hamiltonian ($H_{\sma{op}}$) on an open chain of $L$ sites, to a parafermion Hamiltonian ($\kal$) on a closed ring. For simplicity, we assume that the open-chain Hamiltonian is translational-invariant up to edge corrections, i.e., we can decompose $H_{\sma{op}}=\sum_{r\geq 1}\sum_{j=1}^{L-r+1}V_{j,r}$, where $r$ is the range of the operator $V_{j,r}$, i.e., $V_{j,r}$ acts nontrivially on sites $\{j,j+1,\ldots, j+r-1\}$. Furthermore, we assume $V_{j,r}$ is related to $V_{j+1,r}$ by translation. For a $\Z_3$ example, consider $H_{\sma{op}} = -\sum_{\sma{j=1}}^{\sma{L-1}}\omega \dg{\gamma}_{\sma{2j+1}}\pdg{\gamma}_{\sma{2j}} + h.c.,$ where $V_{j,2} = -\omega \dg{\gamma}_{\sma{2j+1}}\pdg{\gamma}_{\sma{2j}}+h.c.$. For $N>2$, a $\Z_N$ Hamiltonian on a closed chain cannot be uniquely defined by the identification $\gamma_{\sma{2L+j}}=\gamma_{j}$, due to the nontrivial commutation\ (\ref{parafermionalgebra}).\cite{bondesan2013} In our example, we might have tried to extend $H_{\sma{op}}$ by translational symmetry: $H_{\sma{op}} \rightarrow H_{\sma{op}}+ V_{L,2}$, then identified $\gamma_{\sma{2L+1}} \equiv \go$ directly. The result would be a closed-chain Hamiltonian
\begin{align} \label{example1}
\tilde{\calk}^{\sma{(\alpha)}} = -J\sum_{j=1}^{L-1}\omega \dg{\gamma}_{\sma{2j+1}}\pdg{\gamma}_{\sma{2j}} - J\omega^{\alpha-1} \dg{\gamma}_{\sma{1}}\pdg{\gamma}_{\sma{2L}} + h.c.
\end{align}
 with $\alpha=2$. On the other hand, first interchanging $ \dg{\gamma}_{\sma{2L+1}}\pdg{\gamma}_{\sma{2L}} = \omega \pdg{\gamma}_{\sma{2L}}\dg{\gamma}_{\sma{2L+1}}$, and then making the same identification, we arrive at a different Hamiltonian\ (\ref{example1}) with $\alpha=1$. If there exist more than one inter-edge couplings, the phase of each is ambiguous. These ambiguities motivate us to identify $\sigma_{\sma{L+j}}=\sigma_{\sma{j}}$ instead -- since clock operators on different sites commute, the clock Hamiltonian is uniquely defined on a closed chain. Then by attachment of Jordan-Wigner strings we convert this clock Hamiltonian (which is clock-local) to a parafermion Hamiltonian (which is parafermion-local but clock-nonlocal). Let us illustrate this procedure with the family of open-chain Hamiltonians $\hno$ (from Eq.\ (\ref{pfHamn1})), which we extend and then identify $\sigma_{\sma{L+1}}=\sigma_{\sma{1}}$, to obtain clocks on a ring:
\begin{align} \label{dualN1}
\gno =\hno -J \sum_{\beta=0}^{N-1}(\pdg{\sigma}_L \dg{\sigma}_{1})^{\beta}.
\end{align}
In the next step, we decompose the inter-edge coupling into a form $B_{\text{cl}} = B_lB_r$, such that $B_l$ ( resp.\ $B_r$) is a clock-local operator with a definite charge, and with support near site $1$ (resp.\ $L$). To convert $B_{\text{cl}}$ to an inter-edge coupling ($B_{\text{pf}}$) of parafermions, we must attach as many powers of the string $Q$ as the charge of $B_r$. The resultant inter-edge coupling
\begin{align} \label{convert}
B^{\sma{(\alpha)}}_{\text{pf}} = B_lB_r( \omega^{\alpha} Q)^{[[B_r]]}
\end{align}
can always be expressed locally in terms of parafermions, as follows from the Jordan-Wigner transformation\ (\ref{pfclocktransformation}). Here in Eq.\ (\ref{convert}), $[[B_r]]$ refers to the charge of $B_r$, i.e., $\dg{Q}B_rQ = \omega^{[[B_r]]}B_r$, and we have also introduced the twist parameter $\alpha \in \Z_N$. In our example\ (\ref{dualN1}), $[[B_r]]=[[\sigma_L]]=1$, so we attach $\sigma_L \rightarrow \sigma_LQ \sim \pdg{\gamma}_{\sma{2L}}$. Then from $\sigma_L\dg{\sigma}_1Q=\omega^{\sma{(1-N)/2}}\dg{\gamma}_{\sma{1}}\pdg{\gamma}_{\sma{2L}}$, we obtain parafermions-on-a-ring: 
\begin{align} \label{extension}
\kal_{\sma{N,1}} = \hno - J\sum_{\beta=0}^{N-1} \big(\omega^{\alpha+(1-N)/2} \dg{\gamma}_{\sma{1}}\pdg{\gamma}_{\sma{2L}}\big)^{\beta}.
\end{align}

\section{Deriving the topological order parameter in the XYZ model}\label{app:xyztwist}

Our aim is to the derive the topological order parameter for the XYZ Majorana model (\ref{xyzmajorana}), which we parametrize by $J_x$ and $J_y$. There are two parameter regions with distinct order parameters: (i) $|J_x|,|J_y|<1$, as described in App.\ \ref{app:squareC}, and (ii) $|J_y|>1, |J_y|>|J_x|$, described in App.\ \ref{app:regionBD}.  

\subsection{Topological order parameter for $|J_x|,|J_y|<1$} \label{app:squareC}

Let us parametrize the XYZ model (\ref{xyzmajorana}) as
\begin{align}
H_s=H_0+sV,\ins{where} V = \sum_{j=1}^{L-1} \big(\;\gtjmo \gtj \gtjpo \gtjpt -i t \gtjmo \gtjpt\;\big),\;\;s=J_x\;\; t=\frac{J_y}{J_x},
\end{align}
and the frustration-free $H_0$ is defined in Eq.\ (\ref{H0maj}). $H_0$ commutes with the zero edge modes $\go$ and $\gtL$, thus its groundstate space is described by a topological order parameter $\calq=-i\go \gtL$.  If we frustrate this model through $V$, the spectral gap above the lowest two states remain finite in the parameter region: $|J_x|,|J_y|<1$. Our task is to derive the dressed order parameter in this region. We will employ the identities:
\begin{align}
&[V,\go] = -2\gt \gth \gfo + 2i t \gfo  =  (-\gt \gth +it)(w_++w_-), \ins{and} [V,\gtL] = (z_++z_-)(-it+\gtLmt \gtLmo),
\end{align}
where $w_{\pm} = \gfo \pm i \gfi$ and $z_{\pm} = \gtLmth \pm i \gtLmfo$ are eigenoperators under commutation by $H_0$, i.e., $[H_0,w_{\pm}]=\mp 2 w_{\pm},$ and $[H_0,z_{\pm}] = \pm 2 z_{\pm}.$ It follows that
\begin{align}
&i[\cald_0,\go]=\int dt F(2t)e^{iH_0t}[V,\go]e^{-iH_0t} =(it-\gt \gth)(i\gfi) \ins{and} \notag \\
&i[\cald_0,\gtL]=\int dt F(2t)e^{iH_0t}[V,\gtL]e^{-iH_0t} =-i\gtLmfo (-it+\gtLmt \gtLmo),
\end{align}
which determine the edge modes $\calo_l$ and $\calo_r$ to first order in $s$; see Eq.\ (\ref{edgemodefirstorder}). Applying that $\calq=-i\calo_l \calo_r$, we are immediately led to Eq.\ (\ref{TOparameterXYZ}). In this derivation, we have assumed that $|J_x|<1, J_x \neq 0$; if instead $J_x=0$ and $|J_y|<1,J_y \neq 0$, the final answer\ (\ref{TOparameterXYZ}) is identical. 

\subsection{Deriving the topological order parameter for $|J_y|>1, |J_y|>|J_x|$} \label{app:regionBD}

Up to a proportionality constant, we parametrize the XYZ model (\ref{xyzmajorana}) as $H'_s=H'_0+sV'$, where $s=1/J_y$,
\begin{align}
H'_0 =-i \sum_{j=1}^{L-1} \gtjmo \gtjpt \ins{and} V' =  J_x \sum_{j=1}^{L-1} \gtjmo \gtj \gtjpo \gtjpt + i \sum_{j=1}^{L-1} \gtj \gtjpo.
\end{align}
The frustration-free $H_0'$ commutes with the zero edge modes $\gt$ and $\gtLmo$, thus its groundstate space is described by a topological order parameter $\calq^y=-i\gt \gtLmo$, which must be distinguished from the order parameter ($\calq=-i\go \gtL$) of $H_0$.  If we frustrate this model through $V'$, the spectral gap above the lowest two states remain finite in the parameter region: $|J_y|>1, |J_y|>|J_x|$. In this region, the order parameter is dressed as
\begin{align}
\calq^y =  -i \gt \gtLmo + \frac{J_x}{J_y}(\go \gfo \gsix \gtLmo + \gt \gtLmfi \gtLmth \gtL) +\frac{i}{J_y}(\gsix \gtLmo + \gt \gtLmfi) + \ldots
\end{align}
Its derivation straightforwardly generalizes App.\ \ref{app:squareC}. Defining $\bar{\calq}^y$ as the first-order truncation of $\calq^y$ (i.e., dropping the dots), we define a closed-chain Hamiltonian by $H_{\sma{XYZ}}- (-1)^{\alpha}J_y\bar{\calq}^y$, where  $H_{\sma{XYZ}}$ is defined in Eq.\ (\ref{xyzmajorana}). We numerically evaluate the groundstate parities of this closed-chain Hamiltonian for both $\alpha \in \Z_2$, on a chain of $22$ sites. The result is shown in Fig.\ \ref{fig:xyzcomposite}(d); we have whitened the parameter regions where the groundstate parity switches upon flux insertion. Note that $|J_y|>1, |J_y|>|J_x|$ corresponds to regions $B$ and $D$ in this figure.

\section{Finite-size analysis of the translational-invariant Majorana chain} \label{app:duality}

In this Appendix, we present a finite-size analysis of the XYZ Majorana model (\ref{xyzmajorana}) for the parameter region: $|J_x|,|J_y|<1$, which we label as the square $C$ in Fig.\ \ref{fig:xyzcomposite}. Suppose we close the chain with the translational-invariant boundary conditions (TIBC); a detailed description of this procedure is available in App.\ \ref{app:twistmethod}. We then obtain two Hamiltonians, which we denote by $\calk_{\sma{XYZ}}^{\sma{(0)}}$ (resp.\ $\calk_{\sma{XYZ}}^{\sma{(1)}}$) in the case of periodic (resp.\ antiperiodic) boundary condition. A signature of topological order is that the groundstate of $\calkxyz$ is sensitive to its boundary condition; in particular, we expect that the groundstate parity switches between $\calk_{\sma{XYZ}}^{\sma{(0)}}$ and $\calk_{\sma{XYZ}}^{\sma{(1)}}$, for all parameters within the square $C$. A finite-size study shows that the parity is invariant for a subregion (colored black) of $C$ where quantum fluctuations are dominant; see Fig.\ \ref{fig:xyzcomposite}(b). The aim of this Appendix is to explain the effect of finite size on the groundstate sensitivity. Our explanation relies on a Majorana-spin duality which we will first introduce; one motivation for the spin representation is that numerical simulations are often easier with a clock-local basis. \\

\noi{i} To describe this duality, we first relate the open-chain Majorana Hamiltonian ($H_{\sma{XYZ}}$) to a closed-chain spin Hamiltonian ($\calg_{\sma{XYZ}}$) in App.\ \ref{app:twistbyduality}. We argue that $H_{\sma{XYZ}}$ being topologically ordered implies clock-local order in $\calg_{\sma{XYZ}}$.\\

\noi{ii} Then in App.\ \ref{app:dualityxyz}, we show that the closed-chain Majorana Hamiltonian ($\calk_{\sma{XYZ}}^{\sma{(\alpha)}}$) is dual to the closed-chain spin Hamiltonian ($\calg_{\sma{XYZ}}$). Applying this duality, we explain what clock-local order in $\calg_{\sma{XYZ}}$ implies for $\calk_{\sma{XYZ}}^{\sma{(\alpha)}}$. We then analyze the groundstates of $\calkxyzalpha$ and their sensitivity to twisting. \\ 

\noi{iii} Finally in App.\ \ref{app:dualityfinite}, we describe a finite-size comparative study of the three types of boundary conditions discussed in this paper: (i) TIBC, (ii) zeroth-order topological boundary condition (TBC), and (iii) first-order TBC.

\subsection{Relating topological order in $\hxyz$ to clock-local order in the spin Hamiltonian $\calgxyz$} \label{app:twistbyduality}

As defined in Eq.\ (\ref{xyzmajorana}), the open-chain Majorana Hamiltonian can be rewritten as the Heisenberg spin model:
\begin{align}
H_{\sma{XYZ}} = -\sum_{j=1}^{L-1}\sum_{\mu \in \{x,y,z\}} J_{\mu}\varsigma^{\mu}_j\varsigma^{\mu}_{j+1}, \ins{with} J_z=1.
\end{align}
Here, $\varsigma^{\sma{\mu}}_{\sma{j}}$ are Pauli matrices acting on site $j$. To derive the above spin representation, we identify $\varsigma^{\sma{z}} = \sigma$, $\varsigma^{\sma{x}}=\tau$, and apply the Jordan-Wigner transformation\ (\ref{pfclocktransformation}): for bulk couplings ($1\leq  j < L$), 
\begin{align}
\varsigma^x_j\varsigma^x_{j+1} = -\pdg{\gamma}_{\sma{2j-1}}\pdg{\gamma}_{\sma{2j}}\pdg{\gamma}_{\sma{2j+1}}\pdg{\gamma}_{\sma{2j+2}}, \;\; \varsigma^y_j\varsigma^y_{j+1} =i\pdg{\gamma}_{\sma{2j-1}}\pdg{\gamma}_{\sma{2j+2}}, \ins{and} \varsigma^z_j\varsigma^z_{j+1} = -i\pdg{\gamma}_{\sma{2j}}\pdg{\gamma}_{\sma{2j+1}}.
\end{align}
We define $\calg_{\sma{XYZ}}$ as the translational-invariant extension of $H_{\sma{XYZ}}$ to a ring:
\begin{align} \label{Hxyz}
\calg_{\sma{XYZ}} = -\sum_{j=1}^{L}\sum_{\mu \in \{x,y,z\}} J_{\mu}\varsigma^{\mu}_j\varsigma^{\mu}_{j+1}.
\end{align}
Here, periodic spin boundary conditions are imposed as: $\varsigma^{\sma{\mu}}_{\sma{L+1}} = \varsigma^{\sma{\mu}}_{\sma{1}}$. The phase diagram of $\calg_{\sma{XYZ}}$ is well-understood analytically.\cite{Takhtadzhan1979,Elisa} $\calg_{\sma{XYZ}}$ is gapless where $|J_{\mu}|= |J_{\nu}| \geq |J_{\kappa}|$, i.e., for any two parameters having equal magnitude that is greater than or equal to the third magnitude; otherwise, the phase is gapped with a clock-local order parameter $\varsigma^{\mu}$, corresponding to $J_{\mu}$ of the largest magnitude. We illustrate this symmetry-breaking in Fig.\ \ref{fig:duality}(a), for the parameters $(J_x,J_y,J_z)=(0.1,-0.2,1)$. \\

\begin{figure}[H]
\centering
\includegraphics[width=8.3 cm]{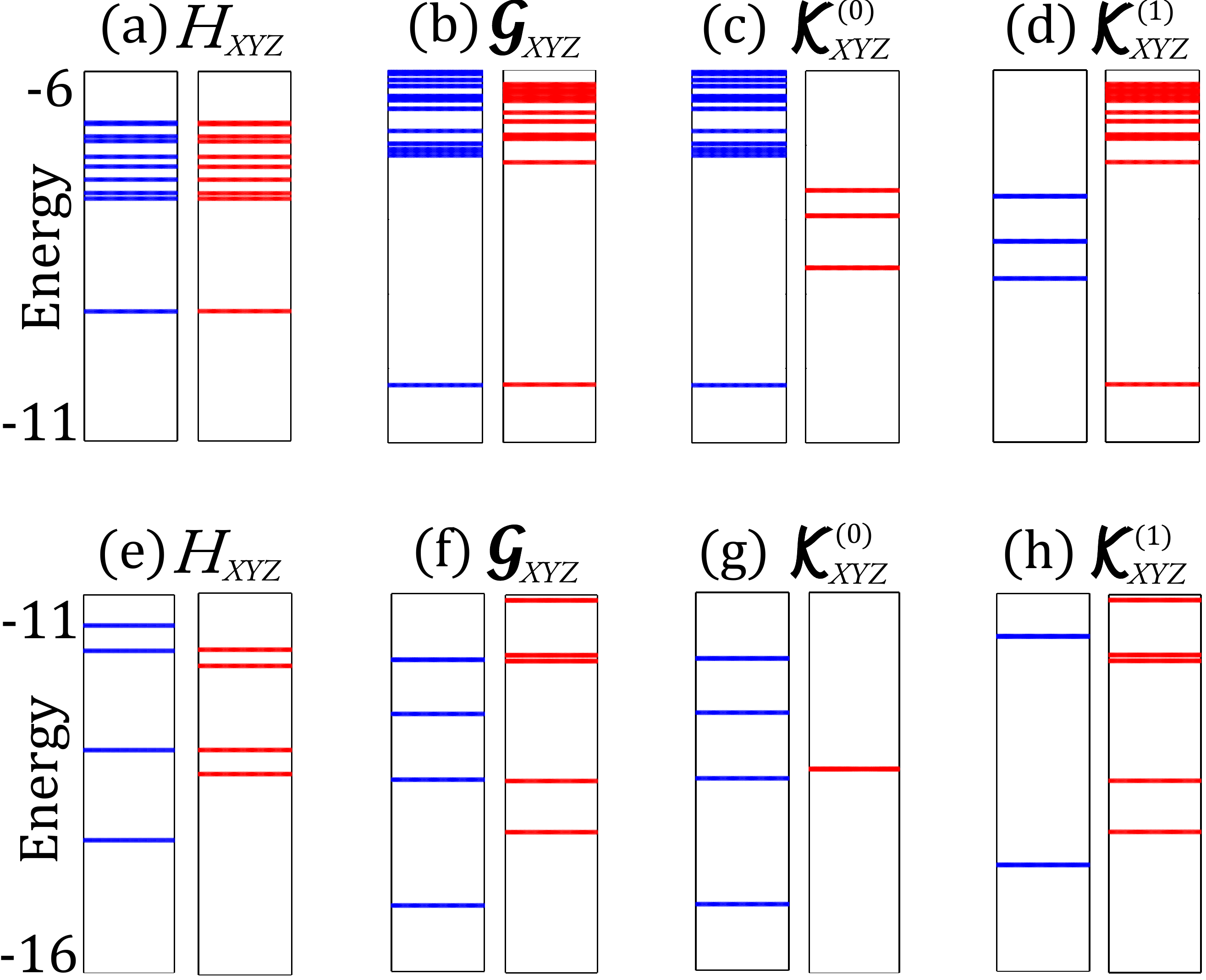}
\caption{ Lowest-lying spectra of the XYZ model on a chain of $10$ sites. Two parametrizations are considered: (a)-(d) correspond to the same parameters  $(J_x,J_y,J_z)=(0.1,-0.2,1)$, while (e)-(h) correspond to $(J_x,J_y,J_z)=(0.75,-0.75,1)$. The spectrum of each Hamiltonian is split into even (blue) and odd (red) charge sectors. (a) and (e): spectrum of the open-chain $H_{\sma{XYZ}}(J_x,J_y,J_z)$, (b) and (f): $\calg_{\sma{XYZ}}$, (c) and (g): $\calk_{\sma{XYZ}}^{\sma{(0)}}$, (d) and (h): $\calk_{\sma{XYZ}}^{\sma{(1)}}$.}  
\label{fig:duality}
\end{figure}

We argue that $\hxyz$ being topologically ordered (in the fermionic language) indicates that $\calgxyz$ is symmetry-broken (in the spin language). This implies that $\hxyz$ and $\calgxyz$ simultaneously manifest a degenerate groundstate manifold, as illustrated in Fig.\ \ref{fig:duality}(a) and (b) for the parameters $(J_x,J_y,J_z)=(0.1,-0.2,1)$. This follows because $\calgxyz$ is the translational-invariant extension of $\hxyz$, thus both Hamiltonians are approximately minimized by the same wavefunctions. Alternatively stated, we expect that the spectral gaps of $\hxyz$ and $\calgxyz$ (above their two lowest-lying states) are related multiplicatively -- a quantum phase transition in $\hxyz$ indicates a simultaneous transition in $\calgxyz$.  To support our hypothesis, we numerically evaluate the spectral gap of $\hxyz$ in Fig.\ \ref{fig:xyzcomposite}(a); the gapless lines (colored black) separate the plot into five regions labelled by $A$ to $E$, with $B,C$ and $D$ corresponding to topological phases (cf. Sec.\ \ref{app:detectTOonring}). Our hypothesis is supported by the exact coincidence of gapless lines in both Hamiltonians ($\hxyz$ and $\calgxyz$). Here, we are comparing our numerical simulation of $\hxyz$ with a known analytical result about $\calgxyz$; cf. the previous paragraph.

\subsection{Majorana-spin duality on a closed chain} \label{app:dualityxyz}

We extend the open-chain $H_{\sma{XYZ}}$ onto a ring through the prescription\ (\ref{convert}), to obtain the closed-chain Hamiltonian $\calk^{\sma{(\alpha)}}_{\sma{XYZ}} = H_{\sma{XYZ}}+\Delta H^{\sma{(\alpha)}}_{\sma{XYZ}}$, as shown in Eq.\ (\ref{xyzinteredgetrans}). The inter-edge coupling can be rewritten as
\begin{align}
\Delta H^{{(\alpha)}}_{\sma{XYZ}} =&\; -J_x \varsigma^x_L\varsigma^x_{1}-J_y\varsigma^y_L\varsigma^y_{1}(-1)^{\alpha}Q -J_z\varsigma^z_L\varsigma^z_{1}(-1)^{\alpha}Q.
\end{align}
Let us define for any Hamiltonian $H$ a projected Hamiltonian $[H]_{\beta}$ in the charge sector $\beta$, i.e., $[H]_{\beta}$ acts in the space of functions with eigenvalue $\omega^{\beta}$ under conjugation by $Q$. Clearly,
\begin{align}
[\Delta H^{{(\alpha)}}_{\sma{XYZ}}]_{\beta} =&\; -J_x \varsigma^x_L\varsigma^x_{1}-J_y\varsigma^y_L\varsigma^y_{1}(-1)^{\alpha+\beta} -J_z\varsigma^z_L\varsigma^z_{1}(-1)^{\alpha+\beta}.
\end{align}
By inspecting the matrix elements of $\hxyz$, $\calgxyz$ and $\calkxyzalpha$ in each charge sector, we derive a duality between Majorana fermions and spins on a closed chain:
\begin{align} \label{dualityxyz}
[\calk^{{(\alpha)}}_{\sma{XYZ}}]_{{\alpha}}= [\calg_{\sma{XYZ}}]_{{\alpha}}, \ins{for} \alpha \in \Z_2.
\end{align}
This duality on a closed chain generalizes an open-chain duality\cite{KitaevLaumann2009,greiter2014} between the transverse-field Ising model and the Kitaev wire. $\calg_{\sma{XYZ}}$ being translational-invariant then also implies the same for $\calk_{\sma{XYZ}}^{\sma{(0)}}$ (resp.\ $\calk_{\sma{XYZ}}^{\sma{(1)}}$) in the even (resp. odd) charge sector, as we alluded to in Sec.\ \ref{sec:transinvBC}.\\ 

This duality is further illustrated in Fig.\ \ref{fig:duality}(b)-(d) for the parameters $(J_x,J_y,J_z)=(0.1,-0.2,1)$. If we define $\ket{\phi^{\sma{J_x,J_y}}_{\sma{\alpha,1}}}$ as the groundstate of $\calgxyz$ in the charge sector $\alpha \in \Z_2$, then clearly 
\begin{align} \label{cond1gua}
\ket{\phi^{\sma{J_x,J_y}}_{{0,1}}} \ins{minimizes} [\calk_{\sma{XYZ}}^{{(0)}}]_0, \ins{and} \ket{\phi^{\sma{J_x,J_y}}_{{1,1}}} \ins{minimizes} [\calk_{\sma{XYZ}}^{{(1)}}]_1.
\end{align}
For $(J_x,J_y,J_z)=(0.1,-0.2,1)$,
\begin{align} \label{cond2ungua}
\ket{\phi^{\sma{J_x,J_y}}_{{0,1}}} \ins{also minimizes} \calk_{\sma{XYZ}}^{{(0)}}, \ins{and} \ket{\phi^{\sma{J_x,J_y}}_{{1,1}}} \ins{also minimizes} \calk_{\sma{XYZ}}^{{(1)}}.
\end{align}
This is illustrated in the spectra of Fig.\ \ref{fig:duality}(b)-(d), and explains why the groundstate parity switches as we twist  $\calk_{\sma{XYZ}}^{\sma{(\alpha)}}$. \\

Due to the duality (\ref{dualityxyz}), the first statement\ (\ref{cond1gua}) is exactly satisfied for any finite size; the second statement\ (\ref{cond2ungua}) is not necessarily satisfied on finite-size chains, i.e., the actual groundstate of $\calk_{\sma{XYZ}}^{\sma{(\alpha)}}$ need not originate from the symmetry-broken manifold of $\calgxyz$. For example, we consider the spectra at $(J_x,J_y,J_z)=(0.75,-0.75,1)$ for a 10-site chain, as plotted in Fig.\ \ref{fig:duality}(e),(f),(g) and (h). Fig.\ \ref{fig:duality}(f) demonstrates the finite-size splitting of the low-energy subspace of $\calg_{\sma{XYZ}}$; in particular, $\ket{\phi^{\sma{3/4,{-3/4}}}_{\sma{0,1}}}$ is energetically favored in comparison with $\ket{\phi^{\sma{3/4,{-3/4}}}_{\sma{1,1}}}$. While $\ket{\phi^{\sma{3/4,{-3/4}}}_{\sma{0,1}}}$ also minimizes $\calk_{\sma{XYZ}}^{\sma{(0)}}$ (see Fig.\ \ref{fig:duality}(g)),  $\ket{\phi^{\sma{3/4,{-3/4}}}_{\sma{1,1}}}$ does not minimize  $\calk_{\sma{XYZ}}^{\sma{(1)}}$ (see Fig.\ \ref{fig:duality}(h)). A corollary is that the parity does not switch, and we have verified that this behavior persists for a chain of $30$ sites. \\

As a side remark, we note that the duality described in Eq.\ (\ref{dualityxyz}) is easily generalized to a parafermion-clock duality: if $\calk^{\sma{(\alpha)}}$ is the Hamiltonian for parafermions on a ring, with a particular twist $\alpha \in \Z_{\sma{N}}$, and $\calg$ is the Hamiltonian for clocks on a ring, then
\begin{align} \label{duality}
[\calk^{\sma{(N-\alpha)}}]_{\alpha}=[{\calg}]_{\alpha}, \ins{with} \alpha \in \Z_N.
\end{align}
Let us exemplify this duality for the frustration-free models $\hno$ defined in Eq.\ (\ref{pfHamn1}). The closed-chain parafermion Hamiltonian\ (\ref{extension}) is dual to the closed-chain clock Hamiltonian ($\gno$) in Eq.\ (\ref{dualN1}). That is, $\calk_{\sma{N,1}}^{\sma{(\alpha)}}$ and $\gno$ satisfy Eq.\ (\ref{duality}), as follows from the identity\ (\ref{examplesupport}).

\subsection{Comparing the three types of closed-chain boundary conditions} \label{app:dualityfinite}

Applying the parity switch as a criterion, we would like to evaluate the various types of closed-chain boundary conditions. We focus on the $J_x=-J_y$ line, for $J_x \in [0,1)$, and define $\bar{J}_x$ such that the parity switches for $0 \leq J_x<\bar{J}_x$, and is invariant for $\bar{J}_x<J_x<1$. With the topological boundary condition (TBC) of Eq.\ (\ref{TBC}), the parity switches for all parameters where the open-chain Hamiltonian is topological, i.e., $\bar{J}_x=1$, as proven in Sec.\ \ref{twist:qc}. With the translational-invariant boundary condition (TIBC) of Eq.\ (\ref{convert}), it has also been argued\cite{zaletel2014} that $\bar{J}_x=1$ for system sizes much larger than the correlation length; in practice this can only be verified with a much larger system size than can be simulated with Lanzcos. In Fig.\ \ref{fig:finitesize}, we numerically evaluate $\bar{J}_x$ for three different types of boundary conditions: (i) TIBC, (ii) zeroth-order TBC (see Sec.\ \ref{twist:qc}), and (iii) first-order TBC (see Eq.\ (\ref{firstTBC})). For all the system sizes that we simulate (up to $L=30$), the groundstates with (i) and (ii) are comparably sensitive to twisting, while the groundstate with (iii) is most sensitive. A naive extrapolation suggests that these conclusions are robust in the thermodynamic limit.

\begin{figure}[H]
\centering
\includegraphics[width=6.6 cm]{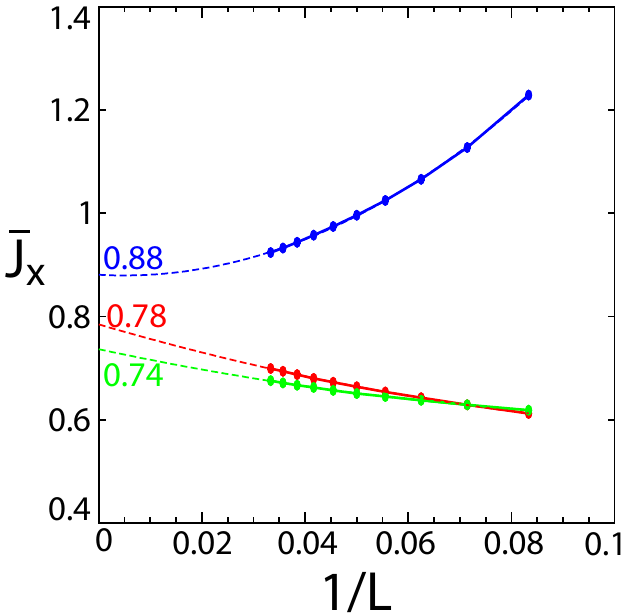}
\caption{ Finite-size analysis of the closed-chain XYZ model along the $J_x=-J_y$ line, for $J_x \in [0,1)$. We close the chain using three types of boundary conditions: (i) translational-invariant (colored red), (ii) first-order topological (blue), and (iii) zeroth-order topological (green). We define $\bar{J}_x$ such that upon twisting the boundary conditions, the groundstate parity switches for $0 \leq J_x<\bar{J}_x$, but is invariant for $\bar{J}_x<J_x<1$. We numerically evaluate $\bar{J}_x$ for varying system sizes (up to $L=30$), as shown by solid dots in the plot. Each dashed line is a fit to the function: $a + b/L + c/L^2$, and by extrapolation we obtain an estimate of $\bar{J}_x$ in the thermodynamic limit. Since the fitting function is not theoretically motivated, we emphasize that these extrapolations are more suggestive than rigorous.}  
\label{fig:finitesize}
\end{figure}

\section{Model with coexisting topological and symmetry-breaking orders} \label{app:modelcoexist}

We present a model example with coexisting topological and symmetry-breaking orders; all other examples in this paper have either been purely-topological or purely-symmetry-breaking. We consider $\Z_8$ parafermions which have four distinct phases labelled by the possible integer divisors: $n \in \{1,2,4,8\}$; each of these phases may be realized by the frustration-free models in Eq.\ (\ref{pfHam}).  Defining $m=8/n$ and $g=$gcd$(m,n)$, we consider the phase labelled by $(n=2,m=4,g=2)$. By the general classification proposed in Ref.\ \onlinecite{bondesan2013}, $g>1$ implies there is symmetry-breaking, and $m>g$ implies topological order; the rest of this Appendix is an elaboration of these statements, with emphasis on concepts we have introduced in this paper, namely their topological edge modes and their local indistinguishability. \\ 
   
We begin with the most intuitive groundstate basis of $\het$ in Eq.\ (\ref{pfHam}) -- these are fully polarized as
\begin{align} \label{82}
&\ket{\psi_{\alpha,2}} = \bigotimes_{j=1}^L\frac{1}{\sqrt{2}}\big(\;\ket{\alpha}_j+\ket{\alpha+4}_j\;), 
\end{align}
with $\alpha \in \Z_4 \equiv \{0,1,2,3\}$ labelling four degenerate groundstates on an open chain. Recall in the above expression that $\sigma_j\ket{\alpha}_j = \omega^{\alpha} \ket{\alpha}_j$. A simple computation shows that $\tau_j^4$ and $\sigma_j^2$ are clock-local order parameters for any subcell coordinate $j$:
\begin{align} \label{clocklocalorder}
\tau_j^4 \ket{\psi_{\alpha,2}} = \ket{\psi_{\alpha,2}} \ins{and} \sigma_j^2\ket{\psi_{\alpha,2}} = \omega^{2\alpha}\ket{\psi_{\alpha,2}}. 
\end{align}
This implies the parafermionic order parameters $\gamma_j^4$ for any $j$, since $\gamma_{2j-1}^4 \propto \sigma_j^4\tau_{j-1}^4 \tau_{j-2}^4 \ldots$ and $\gamma_{2j}^4 \propto \sigma_j^4\tau_j^4\tau_{j-1}^4 \tau_{j-2}^4 \ldots$, through the Jordan-Wigner transformation. While in this model $g=2>1$, we might ask more generally why $g>1$ is necessary for the existence of a parafermionic order parameter. This answer is found in Ref.\ \onlinecite{bondesan2013}, as we summarize:  for the family of Hamiltonians parametrized by $(N,n)$ in Eq.\ (\ref{pfHam}), the generalization of Eq.\ (\ref{clocklocalorder}) is that $\tau^m$ and $\sigma^n$ are clock-local order parameters -- then if there exist integers $a$ and $b$ such that $am=bn$ but is not a multiple of $N$, then one might string together the following order parameter $\sigma_j^{bn}\tau_{j-1}^{am}\tau_{j-2}^{am} \ldots$, which up to a phase factor may be expressed as a parafermionic order parameter: $\gamma_{\sma{2j-1}}^{\sma{cN/g}}$ with $c \in \{0,1,\ldots,g-1\}$; clearly here we need $g>1$.    \\

Returning to our model example, we would like to show that this four-fold degeneracy is really a product of two degeneracies which each arises from a different kind of order -- the topological degeneracy is attributed to edge modes, while the symmetry-broken degeneracy is attributed to the parafermionic order parameter.  One way to proceed is to apply the parafermionic order parameter as a symmetry-breaking `field' -- then if there exists any remnant degeneracy, it must arise from topological order.  As an intermediate step in this computation, we introduce a basis that diagonalizes the $\Z_8$ generator (recall the expression for $Q$ in Eq.\ (\ref{Zngenerator})):
\begin{align} 
\ket{\phi_{\alpha,2}} = \frac{1}{\sqrt{4}}\sum_{\beta=0}^{3}{\omega}^{-2 \alpha \beta} \ket{\psi_{\beta,2}}, 
\end{align}
which satisfies
\begin{align} \label{idenphi}
Q \ket{\phi_{\alpha,2}} = \omega^{2\alpha}\ket{\phi_{\alpha,2}},\;\; \tau_j^4 \ket{\phi_{\alpha,2}} = \ket{\phi_{\alpha,2}} \ins{and} \sigma_j^2\ket{\phi_{\alpha,2}} = \ket{\phi_{\alpha-1 \text{mod} 4,2}}. 
\end{align}
The effect of the symmetry-breaking `field' is to resolve the groundstate manifold as:
\begin{align}
\ket{\theta_{\beta,\delta}} = \frac{1}{\sqrt{2}}\big(\;  \ket{\phi_{\delta,2}} +(-1)^{\beta}\ket{\phi_{\delta+2,2}}\;\big)
\end{align}
with $\beta,\delta \in \Z_2$; this basis splits as
\begin{align}
\gamma_{2j-1}^4\ket{\theta_{\beta,\delta}} =(-1)^{\beta}\ket{\theta_{\beta,\delta}} .
\end{align}
In this sense,  $\beta$ is a symmetry-breaking index that is distinguished by the parafermionic order parameter, while $\delta$ is a topological index that labels each state in each symmetry-broken manifold. While here $\delta \in \Z_2$, in the most general case $\delta \in \Z_{\sma{m/g}}$ and therefore there is no remnant topological degeneracy unless $m>g$. Returning to our model, while $\gamma^4$ breaks the $\Z_8$ symmetry generated by $Q$, there remains a $\Z_4$ symmetry generated by $Q^2$, and a simple computation shows that $\ket{\theta_{\beta,\delta}}$ is an eigenstate of $Q^2$ with eigenvalue $(-1)^{\delta}$. This remnant symmetry generator has the following fractionalized representation in the groundstate space:
\begin{align}
\calq_2 = \omega^2\gamma_1^{-2}\gamma_{2L}^2 = \sigma_1^{-2}\sigma_L^2Q^2.
\end{align}
This operator has the same groundstate matrix elements as $Q^2$, as follows from $\sigma_1^{-2}\sigma_L^2$ acting trivially in the groundstate space. Indeed, from Eq.\ (\ref{idenphi}),  $\sigma_i^{-2}\sigma_j^2\ket{\phi_{\alpha,2}} = \ket{\phi_{\alpha,2}} $ for any $i,j$. Within each symmetry-broken manifold labelled by $\beta$, the topological edge modes $\gamma_1^2$ and $\gamma_{2L}^2$ permute the groundstates as $(\beta,\delta) \rightarrow (\beta,\delta+1 \text{mod} 2)$. Alternatively stated, each symmetry-broken manifold forms a representation of the non-commutative algebra:
\begin{align} \label{noncommutativeex}
\gamma_1^2\gamma_{2L}^2 = &-\gamma_{2L}^2\gamma_1^2, \;\;\;\; \gamma_1^2\calq_2 = -\calq_2\gamma_1^2, \notag \\
 &\text{and}\;\; \gamma_{2L}^2\calq_2 = -\calq_2\gamma_{2L}^2.
\end{align}
While the discussion thus far applies to an open chain, we observe that $\calq_2$ is a legal Hamiltonian term on a closed chain, and its application will split this topological degeneracy.\\

It is instructive to consider a closed-chain Hamiltonian \emph{without} both symmetry-breaking `field' and topological edge modes. Since the `field' is an operator with quartic charge (i.e., $Q^{-1}\gamma_j^4Q = \omega^4\gamma_j^4$), and $\ket{\phi_{\alpha,2}}$ has charge $2\alpha$ (cf. Eq.\ (\ref{idenphi})), we have that the `field' permutes the groundstates as $\ket{\phi_{\alpha,2}} \rightarrow \ket{\phi_{\alpha-2,2}}$. The open-chain manifold then divides naturally into two subspaces which are individually condensates of the `field':
\begin{align}
\bar{P}_{\delta} = \ket{\phi_{\delta,2}}\bra{\phi_{\delta,2}} + \ket{\phi_{2+\delta,2}}\bra{\phi_{2+\delta,2}}, \ins{with} \delta \in \Z_2.
\end{align}
This reveals another perspective to the two different kinds of degeneracies: the topological (symmetry-broken) degeneracy depends (is independent of) the spatial topology; the net degeneracy on an open chain is a product of both degeneracies.

\section{Indistinguishability by symmetric, parafermion-local operators} \label{app:indistpflocal}

Let $\ket{\phi_{\sma{\alpha \in \Z_m,n}}}$ denote the groundstates of the frustration-free Hamiltonian $\hnn$. Our basis is chosen as $Q\ket{\phi_{\alpha,n}}= \omega^{n\alpha}\ket{\phi_{\alpha,n}}$. Assuming $g=\text{gcd}(m,n)>1$, divide $P[\hnn]$ into $(m/g)$ subspaces labelled by $\delta \in \Z_{m/g}$: $\bar{P}_{\delta}=\sum_{\alpha=0}^{g-1}\ket{\phi_{\alpha m/g+\delta,n}}\bra{\phi_{\alpha m/g+\delta,n}}$. Let us use $\bar{\calo}$ to denote a symmetric, parafermion-local probe. We would like to show that the $g \times g$ matrix $\bar{\calm}^{(\delta)}_{\ab} = \bra{\phi_{\alpha m/g+\delta,n}}\bar{\calo}\ket{\phi_{\beta m/g+\delta,n}}$ is proportional to the identity, up to superpolynomially-small finite-size corrections; this is equivalent to the condition\ (\ref{spostrong}). The off-diagonal elements vanish since $[\bar{\calo},Q]=0$. The difference between diagonal elements is
\begin{align} \label{diffdiagg}
\bar{\calm}^{(\delta)}_{\alpha \alpha}- \bar{\calm}^{(\delta)}_{\beta \beta} = \frac{1}{m}\sum_{\mu,\nu=0}^{m-1} \omega^{n\delta(\mu-\nu)}\big(\omega^{\alpha N(\mu-\nu)/g}-\omega^{\beta N(\mu-\nu)/g}\big)\bra{\psi_{\mu,n}}\bar{\calo}\ket{\psi_{\nu,n}},
\end{align}
where $\ket{\psi_{\sma{\mu\in \Z_m,n}}}$ are fully-polarized states\ (\ref{psin}) related to $\ket{\phi}$ by the Fourier transform\ (\ref{chargeeigenbasis}). Crucially, the difference in phase factors vanish when $\mu=\nu$ mod $g$, thus $\bar{\calm}^{(\delta)}_{\alpha \alpha}- \bar{\calm}^{(\delta)}_{\beta \beta}$ reduces to a sum of terms proportional to $ \calt_{\mu,\nu} = \bra{\psi_{\mu,n}}\bar{\calo}\ket{\psi_{\nu,n}}$, with $\mu\neq \nu$ mod $g$. A straightforward generalization of App.\ \ref{app:generalprooflocalindist} shows that $\calt_{\mu,\nu}$ vanishes superpolynomially in the system size, if $\bar{\calo}$ is clock-local. Now we address the case when $\bar{\calo}$ is clock-nonlocal but parafermion-local, as is possible if $\bar{\calo}$ is an inter-edge coupling with support near sites $1$ and $L$. Any inter-edge coupling may be decomposed into the form: $B_lB_rQ^{[[B_r]]}$, where $B_l$ (resp.\ $B_r$) is a clock-local operator with a definite charge, and with support near site $1$ (resp.\ $L$). The nonlocal clock-parafermion transformation\ (\ref{pfclocktransformation}) implies that there are as many powers of the string $Q$ as the charge of $B_r$. Once again, our notation for the charge is: $\dg{Q}B_rQ = \omega^{[[B_r]]}B_r$.  We provide an example of this decomposition for $\Z_4$ parafermions:
\begin{align} \label{z4ex}
\go \gt \gtLmo \gtL = \omega \sigma_1^2 \tau_1 \sigma_L^2 \tau_L Q^2,
\end{align}
where $B_L=\omega \sigma_1^2 \tau_1$, $B_r = \sigma_L^2 \tau_L$ and $[[B_r]]= 2$. \\

We then claim that 
\begin{align} \label{quant}
\calt_{\mu \nu}=\bra{\psi_{\mu,n}}B_lB_rQ^{[[B_r]]}\ket{\psi_{\nu,n}}=0, \ins{for} \mu,\nu \in \Z_m,\;\;\mu \neq \nu \;\text{mod}\;g,
\end{align}
and for finite-ranged $B_r$ and $B_l$; our claim will be proven in App.\ \ref{app:prooflocalindist2}. Eq.\ (\ref{quant}) implies the vanishing of Eq.\ (\ref{diffdiagg}), and thus a proof of Eq.\ (\ref{spostrong}) in the frustration-free limit. If $B_l$ and $B_r$ are superpolynomially decaying into the bulk, Eq.\ (\ref{diffdiagg}) vanishes to up to superpolynomially-small finite-size corrections, which can be bound in a manner analogous to App.\ \ref{app:generalprooflocalindist}. Finally, let us consider deviating symmetrically from the frustration-free limit, while preserving the gap within each of the $g$ charge sectors of $\bar{P}_{\delta}$. Then the deformed subspace is related to the old by a quasi-adiabatic continuation $V_{\delta,s}$, which is distinct for each $g$-multiplet labelled by $\delta$: $\bar{P}_{\delta,s}=V_{\delta,s}\bar{P}_{\delta}\dg{V}_{\delta,s}$. Then applying $[V_{\delta,s},Q]=0$,
\begin{align}
\bar{P}_{\delta,s}\bar{\calo}\bar{P}_{\delta,s} = V_{\delta,s}\bar{P}_{\delta}\dg{V}_{\delta,s}\bar{\calo}V_{\delta,s}\bar{P}_{\delta}\dg{V}_{\delta,s}= V_{\delta,s}\bar{P}_{\delta}(\dg{V}_{\delta,s}B_lV_{\delta,s})(\dg{V}_{\delta,s}B_rV_{\delta,s})Q^{[[B_r]]} \bar{P}_{\delta}\dg{V}_{\delta,s}.
\end{align}
Since $V_{\delta,s}$ is locality-preserving, the dressed operators $\dg{V}_{\delta,s}B_lV_{\delta,s}$ and $\dg{V}_{\delta,s}B_rV_{\delta,s}$ are superpolynomially-decaying into the bulk, thus we may apply our frustration-free result: $\bar{P}_{\delta}(\dg{V}_{\delta,s}B_lV_{\delta,s})(\dg{V}_{\delta,s}B_rV_{\delta,s})Q^{[[B_r]]} \bar{P}_{\delta} \approx c\bar{P}_{\delta}$. The final result is that the indistinguishability condition is robust to superpolynomial accuracy.

\subsection{Vanishing of Eq.\ (\ref{quant}) for finite-ranged $B_l$ and $B_r$} \label{app:prooflocalindist2}

Let us decompose $B_lB_rQ^{[[B_r]]}$ into a sum of terms like $r\prod_{j=1}^L\calo_j$, where $r$ is a nonzero complex number, and $\calo_j$ is an on-site operator acting on site $j$, with the general form $\sigma_j^{a_j}\tau_j^{b_j}$ for integers ${a_j}$ and ${b_j}$. In our $\Z_4$ example\ (\ref{z4ex}), 
\begin{align} \label{caloj}
\calo_j = \begin{cases} \sigma^2_j\tau_j^3 \imp a_j=2, \;\;b_j=3 & \ins{for} j \in \{1,L\} \\
                         \tau_j^2 \imp a_j=0,\;\;b_j=2 &\ins{for} j  \in  \{2,3,\ldots,L-2,L-1\} \end{cases}
\end{align}
We deduce two useful properties of $\{a_j,b_j\}$ from the general form of $B_lB_rQ^{[[B_r]]}$:\\

\noi{a}  $B_r$ has support over a set of sites near $L$, thus we may express $B_r \propto \prod_{\sma{j=L-R+1}}^{\sma{L}} \sigma_j^{a_j}\tau_j^{b_j}$ for some range ($R$) that is much smaller than the length of the chain. Since $[[\sigma_j]]=1$ while $[[\tau_j]]=0$, $[[B_r]] =\sum_{\sma{j=L-R+1}}^{\sma{L}} a_j$. In our example, $B_r = \sigma_L^2\tau_L$ and $[[B_r]]=a_L =2$.\\

\noi{b} $B_l$ and $B_r$ has no support on any site $j$ far away from any edge, where the only nontrivial operation ($\tau_j^{\sma{[[B_r]]}}$) arises due to the string operator $Q^{\sma{[[B_r]]}}$. Thus for $j$ in the bulk, $b_j=[[B_r]]$ and $a_j=0$. In example\ (\ref{caloj}), the bulk sites are labelled by $j \in \{2,3,\ldots,L-2,L-1\}$. \\

Applying Eq.\ (\ref{psin}), we now express Eq.\ (\ref{quant}) as a sum of products of on-site elements, with the form:
\begin{align}
\calt_{\mu \nu}=r \prod_{j=1}^L \bra{\tilde{\mu},n} \sigma_j^{a_j}\tau_j^{b_j}\ket{\tilde{\nu},n}_j, \ins{where} \ket{\tilde{\alpha},n}_j =\frac{1}{\sqrt{n}}\sum_{\beta=0}^{n-1}\ket{\alpha+m\beta}_j,\;\;\sigma_j\ket{\alpha}_j = \omega^{\alpha}\ket{\alpha}_j.
\end{align}
Each on-site element may be evaluated as
\begin{align} \label{eachon}
\bra{\tilde{\mu},n} \sigma_j^{a_j}\tau_j^{b_j} \ket{\tilde{\nu},n}_j =&\; \sum_{\alpha,\beta=0}^{n-1} \frac{\omega^{{a_j}(\nu+{b_j}+m\beta)}}{n}\braket{\mu+m\alpha}{\nu+{b_j}+m\beta} \notag \\
=&\; \sum_{\alpha=0}^{n-1} \frac{\omega^{{a_j}(\mu+m\alpha)}}{n}\delta_{\mu,\nu+{b_j} \,\text{mod}\,m}=  \omega^{a_j\mu}\,\delta_{\mu,\nu+{b_j} \,\text{mod}\,m}\,\delta_{{a_j},0 \,\text{mod}\,n}.
\end{align}
The second equality follows from $\mu$ and $\nu \in \Z_m$. In our $\Z_4$ example, where $g=m=n=2$,
\begin{align} \label{evalon}
&\bra{\tilde{0},2} \calo_j \ket{\tilde{1},2}_j \lin
 \eq \begin{cases}\half (\bra{0}_j+\bra{2}_j) \sigma^2_j\tau_j^3 (\ket{1}_j+\ket{3}_j) =\half (\bra{0}_j+\bra{2}_j) (\ket{0}_j+\ket{2}_j) =1, &\ins{for} j \in \{1,L\} \\
\half (\bra{0}_j+\bra{2}_j) \tau^2_j (\ket{1}_j+\ket{3}_j) =\half (\bra{0}_j+\bra{2}_j) (\ket{1}_j+\ket{3}_j)=0, &\ins{for} j \in \{2,3,\ldots,L-1\}. \end{cases}
\end{align}
From the vanishing of on-site elements in the bulk, we deduce that  $\calt_{01}=0$. Suppose we try to have the bulk terms be non-vanishing; in the following we show that one of the edge terms must consequently vanish! From the second line of Eq.\ (\ref{evalon}), we would need a bulk operator $\tau_j^{b_j}$ with odd $b_j$. From property (b) above, we would then need that $[[B_r]]$ is also odd. This implies from property (a) above that $\sum_{\sma{j=L-R+1}}^{\sma{L}} a_j$ is odd, for $R$ the range of the Hamiltonian, i.e., for at least one site labelled by $\bar{j} \in \{L-R+1,\ldots,L\}$, $a_{\bar{j}}$ is odd. For this particular site, we evaluate
\begin{align}
\bra{\tilde{0},2} \calo_{\bar{j}} \ket{\tilde{1},2}_{\bar{j}} =\half (\bra{0}_{\bar{j}}+\bra{2}_{\bar{j}}) \sigma_{\bar{j}}^{a_{\bar{j}}}\tau^{b_{\bar{j}}}_{\bar{j}} (\ket{1}_{\bar{j}}+\ket{3}_{\bar{j}}). 
\end{align}
If $b_{\bar{j}}$ is even, then this term reduces to a sum of $\braket{0}{1}$, $\braket{0}{3}$, $\braket{2}{1}$, and $\braket{2}{3}$, which individually vanishes. Suppose we choose $b_{\bar{j}}$ to be odd, then
\begin{align}
\bra{\tilde{0},2} \calo_{\bar{j}} \ket{\tilde{1},2}_{\bar{j}} =\half (\bra{0}_{\bar{j}}+\bra{2}_{\bar{j}}) \sigma_{\bar{j}}^{a_{\bar{j}}}(\ket{0}_{\bar{j}}+\ket{2}_{\bar{j}}) = \half (\bra{0}_{\bar{j}}+\bra{2}_{\bar{j}})(\ket{0}_{\bar{j}}+(-1)^{a_{\bar{j}}}\ket{2}_{\bar{j}})=0. 
\end{align}
The last equality follows from $a_{\bar{j}}$ being odd. We have thus proven that one edge term must vanish, and therefore $\calt_{01}=0$ most generally.\\ 

More generally, we would like to show that $\calt_{\mu,\nu} = 0$ for $\mu \neq \nu$ mod $g$. If we assume the converse, we require that each on-site element must be non-vanishing, i.e., from Eq. (\ref{eachon}),
\begin{align} \label{constraintonsite}
\forall j \in \{1,2,\ldots, L\}, \;\; \mu = \nu + b_j \;\text{mod}\;m, \ins{and} a_j = 0 \;\text{mod} \;n.
\end{align}
From our assumption, we now prove a contradiction:\\

\noi{i} For a site $j$ far away from the edge, we derive from Eq.\ (\ref{constraintonsite}) and property (b) that $[[B_r]] = \mu-\nu$ mod $m$. The condition on $\mu$ and $\nu$ then implies that $[[B_r]]$ is \emph{not} an integer multiple of $g$. \\

\noi{ii} $\calt_{\mu,\nu} \neq 0$ implies from Eq.\ (\ref{eachon}) that each of $a_j$ is a multiple of $n$. With property (a), we derive that $[[B_r]] =\sum_{\sma{j=L-R+1}}^{\sma{L}} a_j= nk$,  where $k\in \Z_m$. Since $g=\text{gcd}(m,n)$, this implies $[[B_r]]$ is an integer multiple of $g$, in contradiction with (i). Thus we conclude that $\calt_{\mu,\nu}=0$ for $\mu \neq \nu$ mod $g$.


\end{widetext}

\bibliography{TI-references-2014Aug}

\end{document}